\newif\ifnewColor
\newif\ifsubmission
\newif\ifsmallTable
\newif\ifincludeParams
\newif\ifincludeProbs
\newif\ifincludeOld

\newColorfalse
\submissionfalse
\smallTablefalse
\includeParamsfalse
\includeProbstrue
\includeOldfalse

\ifsubmission
\documentclass[conference]{IEEEtran}
\else
\documentclass[a4paper,11pt,notitlepage]{article}
\usepackage{fullpage}
\fi

\usepackage[USenglish]{babel}
\usepackage[utf8]{inputenc}

\usepackage[T1]{fontenc}
\usepackage{lmodern}
\usepackage{microtype}
\usepackage{bbold}
\usepackage{amssymb}
\usepackage{amsmath}
\usepackage{amsthm}
\usepackage{amsfonts}
\usepackage{nicefrac}
\usepackage{menukeys}
\usepackage{nccmath}
\usepackage{bbm}
\usepackage[b]{esvect}
\usepackage{mathrsfs}
\usepackage{mathtools}
\usepackage{centernot}
\usepackage{xspace}
\usepackage{stmaryrd}
\usepackage{thmtools, thm-restate}
\usepackage{totcount}
\usepackage{color}
\usepackage[most]{tcolorbox} %
\usepackage[colorinlistoftodos]{todonotes}
\usepackage{verbatim}

\usepackage{wrapfig}
\usepackage{graphicx}

\usepackage{float}
\usepackage{caption}
\usepackage{enumerate}
\usepackage{enumitem}
\usepackage{varwidth} %

	\usepackage{authblk}

\usepackage{pgfplots}

\usepackage{algpseudocode}

\usepackage{tikz}
\usetikzlibrary{arrows, positioning}

\usepackage{framed}

\usepackage{makeidx}

\newtheoremstyle{def}%
{}{}%
{}{}%
{\bfseries}{}%
{}{}

\newtheoremstyle{noparens}%
  {}{}%
  {}{}%
  {}
  {}%
  { }%
  {\textbf{\thmname{#1}\thmnumber{ #2}}\thmnote{ #3} \newline}

\theoremstyle{noparens}

\newtheorem{definition}{Definition}

\theoremstyle{remark}

\newtheorem{remark}{Remark}

\theoremstyle{noparens}

\newtheorem{lemma}{Lemma}
\newtheorem{theorem}{Theorem}

\ifsubmission
\def\subAlign{&}
\else
\def\subAlign{}
\fi

\ifsubmission
\def\subBreak{\\}
\else
\def\subBreak{}
\fi

\ifsmallTable
\def\condTable{table}
\else
\def\condTable{table*}
\fi

\ifsubmission
\newcommand{\coqPath}[3]{{\footnotesize \href{\gitlinkext #3}{\menu[,]{#2}}}}
\else
\newcommand{\coqPath}[3]{\href{\gitlinkext #3}{\menu[,]{#1, #2}}}
\fi

\ifsubmission
\newcommand{\coqPathExample}{{\footnotesize \href{\gitlinkNoURL}{\menu[,]{term\_name}} }}
\else
\newcommand{\coqPathExample}{\href{\gitlinkNoURL}{\menu[,]{FileName, term\_name}} }
\fi

\newenvironment{coqdef}[4]
{\begin{definition}[(#1).\hfill \coqPath{#2}{#3}{#4}]}
{\end{definition}}

\newenvironment{coqlem}[4]
{\begin{lemma}[(#1). \hfill \coqPath{#2}{#3}{#4}]}
{\end{lemma}}

\newenvironment{lem}[1]
{\begin{lemma}[(#1).]}
{\end{lemma}}
\newenvironment{coqthm}[4]
{\begin{theorem}[(#1). \hfill \coqPath{#2}{#3}{#4}]}
{\end{theorem}}

\makeindex

\newboolean{stable}
\setboolean{stable}{false}

\newcommand{\hideinstable}[1]{%
\ifthenelse{\boolean{stable}}{}{#1}
}

\algrenewcommand\algorithmicrequire{\textbf{Input:}}

\makeatletter
\algnewcommand{\ExtendedState}[1]{\State
\parbox[t]{\dimexpr\linewidth-\ALG@thistlm}{\hangindent=\algorithmicindent\strut\hangafter=3#1\strut}}
\makeatother

\algnewcommand\algorithmicinput{\textbf{Input:}}
\algnewcommand\Input{\item[\algorithmicinput]}

\algrenewcommand{\algorithmiccomment}[1]{{\color{gray}// #1}}

\algnewcommand{\IIf}[1]{\State\algorithmicif\ #1\ \algorithmicthen}
\algnewcommand{\EndIIf}{\unskip\ \algorithmicend\ \algorithmicif}

 \RequirePackage{color}
 \RequirePackage[most]{tcolorbox}%
 \RequirePackage{totcount}

 \definecolor{blueinfo}{RGB}{64, 112, 173}%
 \definecolor{greeninfo}{RGB}{148, 176, 54}%
 \definecolor{yellowinfo}{RGB}{240, 189, 82}%
 \definecolor{redinfo}{RGB}{194, 77, 59}%
 \definecolor{devpurple}{RGB}{125, 58, 193}%
 \definecolor{implgray}{RGB}{193, 193, 193}
 \definecolor{greencomment}{RGB}{118,191,105}
 \definecolor{bluecomment}{RGB}{105,166,191}
 \definecolor{yellowcomment}{RGB}{239,200,115}
 \definecolor{redcomment}{RGB}{194, 77, 59}
 \definecolor{purplecomment}{RGB}{207, 87, 203}
 \IfFileExists{fontawesome.sty}{%
 \RequirePackage{fontawesome}
 }{%
 \providecommand{\faInfoCircle}{\hspace{3pt}\textbf{i}}
 \providecommand{\faExclamationCircle}{\hspace{3pt}\textbf{!}}
 \providecommand{\faExclamationTriangle}{\hspace{3pt}\textbf{!}}
 \providecommand{\faCogs}{\hspace{2pt}\textbf{D}}
 \providecommand{\faComment}{}
 \providecommand{\faHSquare}{\hspace{2pt}\textbf{H}}
 }

 \newtcolorbox{titlebox}[5]{enhanced,center,colframe=black,colback=white,boxrule={#3},arc={#2},auto outer arc,%
 breakable,pad at break*=5pt,vfill before first,before={\par\medskip\noindent},after={\par\medskip},top=12pt,left=4pt,%
 enlarge top by=7pt,%
 title={\rule[-.3\baselineskip]{0pt}{\baselineskip}\normalsize\sffamily\bfseries #1}, varwidth boxed title*=-30pt,
 attach boxed title to top left={yshift=-10pt,xshift=10pt}, coltitle=black,
 boxed title style={colback=white,boxrule={#5},arc={#4},auto outer arc}
 }

\newtcolorbox{info}{enhanced,colframe=blueinfo,colback=white,boxrule=0.5pt,leftrule=15pt,arc=1pt,auto outer arc,%
 enlarge top by=7pt,%
 overlay={\node[anchor=west,xshift=-1pt,yshift=-13pt,color=white] at (frame.north west) {\faInfoCircle}; }}

\newtcolorbox{notice}{enhanced,colframe=yellowinfo,colback=white,boxrule=0.5pt,leftrule=15pt,arc=1pt,auto outer arc,%
 enlarge top by=7pt,%
 overlay={\node[anchor=west,xshift=-1pt,yshift=-13pt,color=white] at (frame.north west) {\faExclamationCircle}; }}

\newtcolorbox{warning}{enhanced,colframe=redinfo,colback=white,boxrule=0.5pt,leftrule=15pt,arc=1pt,auto outer arc,%
 enlarge top by=7pt,%
 overlay={\node[anchor=west,xshift=-1.5pt,yshift=-13pt,color=white] at (frame.north west) {\faExclamationTriangle}; }}

\newtotcounter{notecount}

\makeatletter
\newtcolorbox[use counter=notecount,list inside=devlist]{@dimplcomment}{%
list entry={\textcolor{implgray}{\faHSquare} \hspace{2pt} Implementation info},%
enhanced,colframe=implgray,colback=white,boxrule=0.5pt,%
leftrule=17pt,arc=1pt,auto outer arc,enlarge top by=7pt,%
overlay={\node[anchor=west,xshift=0pt,yshift=-13pt,color=white] at (frame.north west) {\faHSquare};}%
}
\newcommand{\implinfo}[1]{
\ifthenelse{\boolean{stable}}{}{%
\begin{@dimplcomment}#1\end{@dimplcomment}
}}

\newtcolorbox[use counter=notecount,list inside=devlist]{@devcomment}{%
list entry={\textcolor{devpurple}{\faCogs} \hspace{2pt} Development comment},%
enhanced,colframe=devpurple,colback=white,boxrule=0.5pt,%
leftrule=17pt,arc=1pt,auto outer arc,enlarge top by=7pt,%
overlay={\node[anchor=west,xshift=-1pt,yshift=-13pt,color=white] at (frame.north west) {\faCogs};}%
}
\newcommand{\devcomment}[1]{
\ifthenelse{\boolean{stable}}{}{%
\begin{@devcomment}#1\end{@devcomment}
}}
\newtcolorbox[use counter=notecount,list inside=devlist]{@usercomment}[1]{%
enhanced,colframe=devpurple,colback=white,boxrule=0.5pt,%
arc=1pt,auto outer arc,leftrule=15pt,left=2pt,beforeafter skip=1pt,%
overlay={\node[anchor=west,xshift=-1.5pt,yshift=-13pt,color=white] at (frame.north west) {\faComment};},%
#1}

\newcommand{\bnote}[1]{%
\ifthenelse{\boolean{stable}}{}{%
\begin{@usercomment}{colframe=yellowcomment,list entry={\textcolor{yellowcomment}{\faComment} \hspace{2pt} Bas's comment}}\footnotesize{\textbf{Bas:} #1}\end{@usercomment}
}}
\newcommand{\snote}[1]{%
\ifthenelse{\boolean{stable}}{}{%
\begin{@usercomment}{colframe=purplecomment,list entry={\textcolor{purplecomment}{\faComment} \hspace{2pt} Søren's comment}}\footnotesize{\textbf{Søren:} #1}\end{@usercomment}
}}

\newtcolorbox{bigwarning}{%
enhanced,colframe=devpurple,colback=devpurple!80,%
boxrule=0.5pt,leftrule=30pt,arc=1pt,auto outer arc,%
enlarge top by=7pt,%
overlay={\node[anchor=west,xshift=5pt,yshift=-13pt,color=white] at (frame.north west) {\faCogs};}%
}

\newcommand{\notewarning}{%
\ifthenelse{\boolean{stable}}{}{%
 \ifnum\totvalue{notecount}>0%
  \vspace{1ex}
 \begin{bigwarning}%
   \color{white}{\textbf{The document still contains development comments!}}
  \end{bigwarning}
  \vspace{1ex}
 \fi
}}

\newcommand{\listofdevcomments}{%
  \ifthenelse{\boolean{stable}}{}{%
   \tcblistof{devlist}{List of Development Comments}
}}
\makeatother

\newcommand{\lib}[1]{\textsf{#1}\xspace}
\newcommand{\type}[1]{\textsc{#1}\xspace}
\newcommand{\function}[1]{\texttt{#1}\xspace}
\newcommand{\field}[1]{\function{#1}}

\newcommand{\new}[1]{{\ifnewColor \color{red}\else\fi #1}}
\newcommand{\old}[1]{{\color{blue} \ifincludeOld #1 \else\fi}}

\newcommand{\noprop}{-}
\newcommand{\memeq}{=_{\mathsf{i}}}
\newcommand{\Ex}{\mathbb{E}} %
\newcommand{\aquad}{\mkern9mu}

\newcommand{\rvarLS}{\mathsf{LS}} 
\newcommand{\rvarSS}{\mathsf{SS}} 
\newcommand{\rvarAS}{\mathsf{AS}}

\newcommand{\pLS}{p_{\mathsf{LS}}}
\newcommand{\pSS}{p_{\mathsf{SS}}}
\newcommand{\pAS}{p_{\mathsf{AS}}}
\newcommand{\nat}{\mathbb{N}} 
\newcommand{\bool}{\mathbb{B}} 

\newcommand{\option}{\type{option}} 
\newcommand{\Unit}{\type{Unit}} 
\newcommand{\Type}{\type{Type}} 
\newcommand{\finType}{\type{finType}} 
\newcommand{\eqType}{\type{eqType}} 

\newcommand{\Party}{\type{Party}} 
\newcommand{\Transactions}{\type{Txs}} 

\newcommand{\LocalState}{\type{LocalState}} 
\newcommand{\GlobalState}{\type{GlobalState}} 

\newcommand{\worldv}{N}         %

\newcommand{\AdversarialState}{\type{AdversarialState}} 
\newcommand{\AdvStateI}{\function{AdversarialState}_{0}} 

\newcommand{\treeType}{\type{treeType}} 
\newcommand{\tTT}{\type{tT}}
\newcommand{\Tree}{\type{Tree}} 
\newcommand{\Slot}{\type{Slot}} 
\newcommand{\Message}{\type{Message}} 
\newcommand{\Messages}{\type{Messages}} 

\newcommand{\MsgTuple}{\type{MsgTuple}} 
\newcommand{\MsgTuples}{\type{MsgTuples}} 
\newcommand{\Delay}{\type{Delay}} 
\newcommand{\DelayMap}{\type{DelayMap}} 

\newcommand{\Progress}{\type{Progress}} 

\newcommand{\Hash}{\type{Hash}} 
\newcommand{\Block}{\type{Block}} 
\newcommand{\Chain}{\type{Chain}} 
\newcommand{\seq}{\type{seq}} 

\newcommand{\tv}{\type{T}}      %
\newcommand{\ROState}{\type{OracleState}} 
\newcommand{\id}{\field{id}}
\newcommand{\tT}{\field{tT}}
\newcommand{\tree}{\field{tree}}
\newcommand{\pred}{\field{pred}}
\newcommand{\bid}{\field{bid}}
\newcommand{\msg}{\field{msg}}
\newcommand{\rcv}{\field{rcv}}
\newcommand{\cd}{\field{cd}}
\newcommand{\txs}{\field{txs}}
\newcommand{\honestRcv}{\function{honest\_rcv}}
\newcommand{\honestBake}{\function{honest\_bake}}
\newcommand{\AdversarialBake}{\function{AdversarialBake}}
\newcommand{\AdversarialRcv}{\function{AdversarialRcv}}
\newcommand{\Winner}{\function{Winner}}
\newcommand{\TotalWinner}{\function{TotalWinner}}
\newcommand{\HashBlock}{\function{HashBlock}}
\newcommand{\Honest}{\function{Honest}}
\newcommand{\GenesisBlock}{\function{GenesisBlock}}
\newcommand{\TxSelection}{\function{TxSelection}} 

\newcommand{\RO}{\function{RO}}
\newcommand{\NI}{\function{N}_{0}}

\newcommand{\InitParties}{\function{InitParties}}
\newcommand{\luckySlot}{\function{lucky\_slot}}

\newcommand{\vc}{\function{valid\_chain}}
\newcommand{\slot}{\function{slot}}
\newcommand{\BlockMsg}{\function{BlockMsg}}

\newcommand{\floodMsgs}{\function{flood\_msgs}}
\newcommand{\floodMsgsAdv}{\function{flood\_msgs\_adv}}
\newcommand{\Ready}{\function{Ready}}
\newcommand{\Delivered}{\function{Delivered}}
\newcommand{\Baked}{\function{Baked}}

\newcommand{\emptyseq}{\left[:: \right]}
\newcommand{\cat}{+\!\!\!\!+}

\newcommand{\TreeTypeMap}{\function{TreeTypeMap}} %

\newcommand{\treeI}{\function{tree}_{0}} %
\newcommand{\extendTree}{\function{extendTree}} 
\newcommand{\allBlocks}{\function{allBlocks}} 
\newcommand{\bestChain}{\function{bestChain}} 

\newcommand{\pos}{\function{pos}} 
\newcommand{\cfb}{\function{cfb}} 
\newcommand{\prune}{\function{prune}} 
\newcommand{\suffix}{\preceq} 

\newcommand{\mathcomp}{\lib{mathcomp}}
\newcommand{\finmap}{\lib{finmap}}
\ifsubmission
\newcommand{\coq}{Coq\xspace}
\else
\newcommand{\coq}{\lib{Coq}}
\fi
\newcommand{\isabelle}{Isabelle\xspace}
\newcommand{\equations}{\lib{coq-equations}}
\newcommand{\ocaml}{\lib{OCaml}}
\newcommand{\toychain}{Toychain\xspace}
\newcommand{\easycrypt}{Easycrypt\xspace}

\newcommand{\NatPair}{\type{NatPair}}
\newcommand{\aField}{\field{a}}
\newcommand{\bField}{\field{b}}

\usepackage[colorlinks,allcolors=blue!70!black]{hyperref} %

\usepackage[noabbrev, capitalise]{cleveref} %

\crefname{event}{Event}{Events}             
\newcommand{\Title}{Formalizing Nakamoto-Style Proof of Stake}
\ifsubmission
\title{\Title}
\else
\title{\textbf{\Title}}
\fi
\ifsubmission
\author[]{Søren Eller Thomsen}
\author[]{Bas Spitters}
\affil[]{Concordium Blockchain Research Center, Aarhus University, Denmark\authorcr%
  \{\href{mailto:sethomsen@cs.au.dk}{\nolinkurl{sethomsen}}, \href{mailto:spitters@cs.au.dk}{\nolinkurl{spitters}}\}\nolinkurl{@cs.au.dk}}

\else
\author[]{Søren Eller Thomsen}
\author[]{Bas Spitters}
\affil[]{Concordium Blockchain Research Center, Aarhus University, Denmark\authorcr%
  \{\href{mailto:sethomsen@cs.au.dk}{\nolinkurl{sethomsen}}, \href{mailto:spitters@cs.au.dk}{\nolinkurl{spitters}}\}\nolinkurl{@cs.au.dk}}
\fi

\ifsubmission
\newcommand{\gitlinkNoURL}{https://github.com/AU-COBRA/PoS-NSB}
\newcommand{\gitlink}{\url{\gitlinkNoURL}}
\newcommand{\gitlinkext}{https://github.com/AU-COBRA/PoS-NSB/blob/8cb62e382f17626150a4b75e44af4d270474d3e7/}

\else
\newcommand{\gitlinkNoURL}{https://github.com/AU-COBRA/PoS-NSB}
\newcommand{\gitlink}{\url{\gitlinkNoURL}}
\newcommand{\gitlinkext}{https://github.com/AU-COBRA/PoS-NSB/blob/8cb62e382f17626150a4b75e44af4d270474d3e7/}
\fi

\begin{document}

\maketitle
\begin{abstract}
  Fault-tolerant distributed systems move the trust in a single party to a majority of parties participating in the protocol.
  This makes blockchain based crypto-currencies possible: they allow parties to agree on a total order of transactions without a trusted third party. 
  To trust a distributed system, the security of the protocol and the correctness of the implementation must be indisputable.

  We present the \emph{first} machine checked proof that guarantees both safety and liveness for a consensus algorithm. We verify a Proof of Stake (PoS) Nakamoto-style blockchain (NSB) protocol, using the foundational proof assistant \coq.
  In particular, we consider a PoS NSB in a synchronous network with a static set of corrupted parties. We define execution semantics for this setting and prove chain growth, chain quality, and common prefix which together imply both safety and liveness.
\end{abstract}

\notewarning

\section{Introduction}
\label{sec:intro}
A Byzantine Agreement~\cite{ba82} (BA) protocol allows a group to agree on a decision, even when some of its members behave dishonestly.
Such a protocol is required to satisfy
\begin{description}[leftmargin = \parindent, labelindent= \parindent]
\item[Safety] all honest parties reach the same decision;
\item[Liveness] a decision is reached eventually.
\end{description}
This problem naturally extends to agreeing multiple times (multi-shot-consensus or just consensus).
Until 2008, the main algorithmic approach for achieving consensus was to collect a majority of votes on a decision before taking the next decision.
We will refer to protocols based on this design as quorum-based protocols.

In 2008 Nakamoto's Bitcoin protocol~\cite{nakamoto09} revolutionized the field by introducing a fundamentally different approach for solving the problem.
Instead of letting parties agree on each step of progress by multiple rounds of communication between them, Nakamoto introduced a simple protocol where parties probabilistically take turns making individual progress and disseminating this to all other parties.
If parties often enough have time to see what other parties have disseminated before they make progress, this protocol guarantees safety and liveness up to a negligible probability of failure.

The protocol works by letting all parties maintain an order-preserving data-structure over previous decisions (a block tree) and run a ``lottery'' to decide who is allowed to append the next block to an existing chain in the block tree.
Whenever there is a winner of the lottery, they produce a block and disseminate it to all other parties.
Parties receiving a block will perform a series of checks to guarantee that the block is valid and that the party that produced the block actually won the lottery; if all the checks are correct, the parties should append the new block to their local block tree.
A party will consider a slightly pruned version of their current longest valid chain to be the ordering of blocks agreed upon.
We call a protocol with a similar shape, regardless of lottery mechanism, a Nakamoto-style Blockchain (NSB).

\new{For an NSB there are three main properties that together ensure both liveness and safety~\cite{bbp15}.
  These are \emph{chain growth}, \emph{chain quality} and \emph{common prefix}.
  Chain growth says that the length of the best chain of an honest party increases over time.
  Chain quality says that within a sufficiently large consecutive chunk of blocks of a best chain some of them must be honest.
  Common prefix says that the best chains of honest parties will be a prefix of each other if we remove some blocks from the chain.
}

Because parties probabilistically make individual progress without waiting for a quorum, the lottery needs to be configured in such a way that the time between winners of the lottery must be long enough for blocks to propagate between parties.
NSBs are therefore \emph{only} secure in a \emph{synchronous network}~\cite{DLS88}, where an upper bound on the time it takes to deliver a message is known.
Traditional quorum-based algorithms can be designed such that they are secure in either a synchronous network or a partially-synchronous network~\cite{DLS88} where there exists an unknown upper bound on message delivery time.
The latter requires stronger honesty assumptions.
\\

Nakamoto's original protocol was based on a lottery that assumes that the majority of the computing power participating in the protocol behaves honestly.
The lottery functions by requiring that for a message (block) to be considered valid, the hash of the message needs to be less than a certain threshold.
To participate in the lottery parties will, therefore, try to append different numbers to the messages they want to send, until they find a number which when appended to the message gives a hash which is less than the threshold.
Such a lottery is called a Proof of Work lottery (PoW).
Unfortunately, this design comes with a high power consumption to provide a secure protocol, as honest parties need to ``mine'' more valid messages than dishonest parties to ensure safety.
This problem is solved by the introduction of a Proof of Stake (PoS) lottery, where parties instead can prove that they have the right to create a message for a particular round with their signature.
This construction requires that the majority of stake (for some deterministic calculation of ``stake'') in the system behaves honestly.\\

Because consensus protocols are distributed, they are notoriously difficult to prove correct.
In fact, some protocols were claimed to be both safe and live and passed peer review, but were later found to be either only safe or only live~\cite{bft17}.

Our work establishes both safety and liveness of a PoS NSB.
To make our proofs indisputable we model a PoS NSB protocol with an abstract lottery, provide precise execution semantics for this, and reduce our proofs of safety and liveness for this protocol all the way to the axioms of mathematics using the \coq proof assistant~\cite{the_coq_development_team_2020_3744225}. 
The formalization can be found at
\begin{quote}
\gitlink.
\end{quote}
The formalization uses \coq 8.11.2 with \mathcomp 1.11.0~\cite{ssr10}, \finmap 1.5.0 and \equations 1.2.2~\cite{eq19}. The mathematical components (\mathcomp) library has been used to formalize large parts of mathematics.
It introduces a particular proof style that scales well to large developments and revolves around small-scale-reflection, which we also use for this formalization.

\subsection{Contributions}
\label{sec:contributions}
We implement the behavior of honest parties participating in a PoS NSB with an abstract lottery in \coq.
We use this to define the semantics of the execution of the protocol in a synchronous network accounting for the case when a static subset of parties behaves dishonestly.
We prove that the protocol is \emph{both} safe and live assuming appropriate conditions on the hash-function and the lottery, and restrictions on an adversary's capability to produce honest signatures.
\new{We use the methodology of abstract specification from programming languages.
  This allows us to focus on the core combinatorial arguments that are used in the theory of secure distributed systems.
  To enable this we make some simplifications about signature-schemes that are reminiscent of symbolic cryptography.
  However, our analysis is not a consequence of a series of rewrite rules, but instead we leverage \coq to discover and generalize non-trivial induction-invariants.
}

In particular, we contribute with the following:
\begin{enumerate}
\item We provide the first formalization of \emph{any} consensus algorithm that ensures both safety and liveness in a Byzantine setting.
  Specifically, we verify that a PoS NSB protocol with an abstract lottery \new{and a symbolic signature scheme} ensures consensus.
  In order to do so, we provide precise semantics for executions of distributed protocols with statically corrupted parties in a \emph{synchronous network}.
  \new{As we treat the lottery and signature scheme abstractly, we do not achieve computational security guarantees, but instead focus on the combinatorial arguments which is common in distributed systems.
    We use the semantics to formally prove both chain quality, chain growth and common prefix.
    Our theorems for chain growth and chain quality only requires an honest majority ($>\frac{1}{2}$) of stake which matches the bounds of earlier non-mechanized proofs whereas, for common prefix, our proof requires an honest super-majority ($> \frac{2}{3}$).}

\item \new{In addition to the formalization, we develop a methodology for verifying protocols by abstract functional interfaces, rather than specific non-optimized implementations.
    This may seem to increase the gap between our formalized proof and a running implementation.
    However, by using a precise abstract interface we clearly distinguish the correctness of performant code and that of the protocol.
    We only focus on the latter and isolate the core combinatorial arguments.
    As a side benefit our proof also works for a protocol that allows participants to run different concrete implementations of this abstract interface.
    This is a realistic scenario for a blockchain protocol where different parties might participate with different devices.
    This methodology applies both to pen-and-paper proofs as well as formalizations.}
\end{enumerate}

\subsection{State of the Art}
\label{sec:state-of-art}
To provide context for this work, we give an overview of the state of the art.
First, we provide an overview of analysis of NSBs and next an overview of existing mechanized proofs for consensus algorithms.
In \cref{sec:related-work} we provide a broader comparison to other related work.

\subsubsection{NSB Analysis}
The first cryptographic analysis of a PoW NSB~\cite{bbp15} proved that the protocol underlying Bitcoin satisfies both safety and liveness.
In order to do so they introduced the properties chain quality, chain growth and common prefix, which together imply both safety and liveness.
Their foundational analysis has been extended in several directions: The security has been analyzed in the UC-model~\cite{bcuc17} and the analysis has been modified to cover variations of how the best chain is selected with improved properties~\cite{weight20}. Ren~\cite{ren19} simplifies the original analysis.

In a PoW lottery, a winning event is tied to a specific block, which means that only the particular block that with a hash lower than the threshold will be considered valid by honest players.
In PoS, however, a winning event corresponds to a party being able to sign a block that will be considered valid, which means that nothing prevents an adversary from signing multiple different blocks.
Due to this attack vector a PoS protocol is inherently more difficult to analyze.

The first analysis made for a PoS NSB, was for a lottery with a unique winner in each round~\cite{our17}, which was followed up by an analysis of a lottery that allowed for multiple winners in each round and was generalized to a weaker network model~\cite{op18}.
Similar analysis have later been performed in a composable framework~\cite{og18} and the bounds have been improved~\cite{linc20}.

This work formalizes an analysis similar to previous PoW analysis, but adapts these to work for a PoS lottery.
Our proof roughly follows the proofs in~\cite{weight20}, which in order to analyze different rules for selecting the best chain rule, stated their analysis with a clear separation of necessary conditions on the lottery and combinatorial arguments.
\new{The main difference between our proof and theirs is in the proof of the common prefix property. This argument is quite different for PoS than for PoW.}
Our proof revolves around the fact that the block corresponding to an adversarial lottery ticket can appear at most once on each chain, whereas their proof revolved around that an adversarial block can appear at most once across all chains.
This implies that our proof for common prefix requires $\frac{2}{3}$ of the stake to be honest.
A $\frac{1}{2}$ honesty bound can be obtained for PoS protocols~\cite{our17, op18, linc20} by more complicated proofs revolving around the notion of \emph{characteristic strings}.

\subsubsection{Formalization of Consensus Protocols}
\cref{tab:rw-overview} provides an overview of selected previous formalizations of consensus algorithms in \coq. 

\begin{\condTable}[t]
  \centering
  \ifsmallTable
  \resizebox{\columnwidth}{!}{
    \else\fi
    \begin{tabular}{|l|l|\ifsmallTable\else l | \fi c|c|}
      \hline
      \textbf{Formalization} & \textbf{Type} & \ifsmallTable\else \textbf{Network} & \fi \textbf{Safety} & \textbf{Liveness} \\
      \hline
      \hline    
      \toychain~\cite{toychain18} & PoW NSB &  \ifsmallTable\else Partially synchronous & \fi $(\noprop)$ & $\noprop$ \\
      \hline
      Velasarios~\cite{velisarios18} & Quorum-based & \ifsmallTable\else Partially synchronous & \fi $\checkmark$ & $\noprop$ \\
      \hline    
      Algorand~\cite{algorand19} & Quorum-based & \ifsmallTable\else Partially synchronous & \fi $\checkmark$ & $\noprop$ \\
      \hline    
      Gasper~\cite{gasper20} & Quorum-based & \ifsmallTable\else No execution semantics & \fi $\checkmark$ & $(\noprop)$ \\
      \hline    
      This work & PoS NSB & \ifsmallTable\else Synchronous & \fi $\checkmark$ & $\checkmark$ \\
      \hline
    \end{tabular}
    \ifsmallTable
  }
  \else\fi
  \caption{Overview of previous formalizations in \coq.
    \ifsmallTable\else The formalization of Gasper does not provide execution semantics for the protocol, and so no network-model appears in their formalization. \fi
    \new{By $(\noprop)$ we indicate that only very weak results has been proven about the property.
    In particular, \cite{toychain18} only proves functional correctness and \cite{gasper20} only proves plausible liveness.}
  }
  \label{tab:rw-overview}
\end{\condTable}

\paragraph{Formalization of NSBs}
\toychain~\cite{toychain18} was the first verification effort towards formal guarantees for any NSB (in particular a PoW NSB).
They defined a relation on global states and proved basic properties about the reachable global states.
In a partially synchronous network, they proved that if the system ends up in a state where no messages are waiting to be delivered, then all clients agree on the current best chain.
Although that is an important property of the system it is not enough to argue about how the tree of blocks evolves when the protocol is run, as it will probably never be the case that there are no messages in transit (messages sent but not yet delivered). \toychain did not consider any Byzantine behavior and only focused on functional correctness.

Our work takes the same approach as taken in \toychain, by defining a relation on reachable global states and proving properties for these reachable states. We do, however, model a synchronous network instead of a partially synchronous one, in which stronger properties hold.

\toychain has been extracted and connected to \ocaml-code~\cite{toychainThesis}, which provides an executable node with formal guarantees.
Kaizen~\cite{kaizen19} extends the statements proven in~\cite{toychain18} to apply for an actual performant implementation of a NSB through a series of refinements and transformations of the original code-base, at the cost of a slightly larger trusted computing base.
This work does, however, not improve on the statements proven in \cite{toychain18}.

Probchain~\cite{probchain18} aims to formalize the analysis from~\cite{bbp15}, but they state that their proofs are unfinished.

\paragraph{Formalization of quorum-based consensus}
Traditional (quorum-based) Byzantine fault-tolerant (BFT) consensus algorithms are also used for blockchains.
Velisarios~\cite{velisarios18} is a general framework for formally proving quorum-based BFT algorithms secure in \coq.
They prove a safety property of a widely used BFT algorithm, PBFT~\cite{pbft99}, but do not prove liveness.

A formalization of the Algorand consensus protocol~\cite{algorand19} verifies safety of their BFT algorithm.
Their proof revolves around a transition relation on global states, which models a partially synchronous execution of the protocol.

Ethereum is planning to use a BFT algorithm as a finality layer.
The Casper finality layer has been formally proven to achieve its safety property~\cite{cbc-casper19} in the \isabelle proof assistant.
In \coq, Casper has been proven to be both safe and \emph{plausible live}~\cite{casper18}.
Plausible live is a weaker form of liveness that ensures the protocol will never deadlock.
This result was extended to also cover the revised protocol Gasper which works with a dynamic set of validators~\cite{gasper20}.
The results are proven with an abstract model of quorums on a set of messages without explicitly defining honest behavior and communication.

\subsection{Paper Outline}\label{sec:paper-outline}
The remainder of the paper is organized as follows.
\cref{sec:prelim} describes our notation and conventions.
\cref{sec:protocol} describes how a PoS NSB functions.
In \cref{sec:formal-model} we will introduce the formal setting for our protocol, present the requirements for an implementation of a blocktree, define honest and adversarial behavior, and finally define \emph{reachable} global states.
\cref{sec:results} will present our general results including both the formal theorems and intuition behind the formal proofs.
\cref{sec:related-work} will relate this work to previous work on formalizing distributed systems.
Finally, \cref{sec:concl-future-work} concludes.

\section{Notation}\label{sec:prelim}
The set of natural numbers is denoted $\nat = \{0,1,2,\dots\}$ and boolean values are denoted $\bool = \{\top, \bot\}$.
We adopt conventions from \mathcomp and let $\eqType$ be a type with decidable equality and $\finType$ be a type with a finite duplicate free enumeration.

A record type with the fields $\aField$ and $\bField$ of type $\nat$ is defined by $\NatPair \coloneqq \{\aField : \nat, \bField : \nat \}$.

$\seq\ T$ is the type of lists of type $T$.
$\emptyseq$ denotes the empty sequence, $\left[:: x\right]$ the list with the single element $x$ and $\cat$ the concatenation operator.
We overload standard set notation for filtering and cardinality of sets to also apply to sequences.
\new{We adopt notation from \mathcomp.
  We write $\memeq$ to denote that two sequences have the same members.
  We write $s_{1} \subseteq s_{2}$ to denote that each member in the sequence $s_{1}$ also in $s_{2}$.}

We will use \texttt{teletypefont} for functions and variable names and \textsc{small capitals} for types.

CamelCase (capitalized) names are used for parameters of the formalization and types whereas snake\_case is used for constructs explicitly defined within the formalization.

\coqPathExample are clickable links that directs to the formal definition of the described concept.
\section{The Protocol}\label{sec:protocol}
We consider a static stake PoS NSB protocol similar to the one in~\cite{op18}.
This section provides an informal description of the protocol, such that the description of the formal model and the exact behavior of honest parties presented in~\cref{sec:formal-model} can be guided by intuition.
\\

We discretize time into \emph{slots} which we assume to be totally ordered: $\Slot \triangleq \nat$.
Each party has access to a clock they can query for the current slot, a flooding network they can use to flood messages to each other, and a lottery functionality they can query to check if they are the winner of a slot. We say that a party that wins the lottery for a slot is a \emph{baker} of this slot.
Blocks contain a slot number, a hash of the predecessor, a identifier of the baker, and a signature.
These are the content of messages send through the flooding network in the protocol.

Each party maintains a block tree that initially only contains a single block called the \emph{Genesis Block}. When a block $b$ is added to the block tree it will be added as a successor to the block in the tree with a hash that matches the predecessor of $b$. A path originating at the Genesis Block in a block tree is called a chain. 

The protocol proceeds in slots where each party will do the following for a slot:
\begin{enumerate}
\item Collect all previous blocks that they have received since the last round through the flooding network and add these to their block tree if the signature is valid and the identifier of the block corresponds to a winning party for the round.
\item Evaluate the lottery to check if they are a winner of this round. If they win this slot they will:
  \begin{enumerate}
  \item Calculate what their current longest chain is (disregarding blocks with a higher slot number than the current slot)\footnote{Adversarial parties might choose to evaluate the lottery ahead of time and send these to honest parties.}.
    If there are multiple longest chains of equal length they will use a tie-breaker of their choice to determine the one they consider the best\footnote{This tie-breaker is insignificant for the security of the protocol.}.
  \item Create a new block that will include a hash to the head of their best chain, the current slot, their identity, and their signature. 
  \item Flood this new block using the flooding network.
  \end{enumerate}
\end{enumerate}

The protocol ensures that the participants of the protocol will agree on the current longest chain when removing a few blocks from the head of this chain.
It is for the chains calculated in this way we wish to ensure both safety and liveness.
Specifically, we want to ensure that the best chain of any party grows (chain growth), that honest blocks regularly are appended to this chain (chain quality), and that this chain is both consistent among parties and persistent when the protocol progresses (common prefix).

\section{Formal Model}
\label{sec:formal-model}

We model the protocol described in~\cref{sec:protocol} in a synchronous network with a static but active adversary.
This section describes in detail how this translates to the formal setting in which we prove our results.
First, we present the basic constructs and parameters of our protocol.
Next, we introduce the abstraction and specification of the block tree.
Then we move on to describe our specification of the actual protocol i.e., how honest parties should behave, the global state of the entire system, and the formalization of the synchronous network.
Finally, we put this together to define a relation on what state are reachable from the initial state when running the protocol with a fixed set of parties and a static but active adversary.

We will use this definition of reachable states extensively in~\cref{sec:results}, as we quantify all of our main statements over reachable states.

\subsection{Parameters and Basic Constructs}
\label{sec:parameters}
\new{Our model is parameterized by a type $\Party : \finType$ that represents a unique identifier for a party\footnote{We make this a finite type as there as a finite supply of IP-addresses.}, an equality type $\Transactions : \eqType$ that represents transactions (i.e., content that can be put on the blockchain), and a type $\Hash:\eqType$ that represents the co-domain of a hash function for blocks, $\HashBlock$.} %
A block, $\Block$, is defined to be a record containing four fields
\new{\[ \{ \pred : \Hash,\ \slot : \Slot,\ \txs : \Transactions,\ \bid : \Party \}.\]}
\new{A block contains the predecessor of the block, $\pred$, a slot number in which the block was created, $\slot$, some transactions, $\txs$, and a baker-identifier, $\bid$.}
A \emph{chain} is a sequence of blocks $\Chain \triangleq \seq\ \Block$.

\paragraph{Lottery}
Our model is further parameterized by a predicate, $ \Winner : \Party \rightarrow \Slot \rightarrow \bool$, that allows to check if a particular party has the right to create a block in a specific slot.
\new{This abstraction is intended to capture a lottery similar to the one proposed in the static-stake version of Ouroboros Praos~\cite{op18}.
  There it is determined whether a party wins by evaluating a verifiable random function (VRF) on the current slot number and compare it to a threshold depending on that party's stake.}
\new{We do not model that only persons knowing the secret key can evaluate the lottery.
  Neither do we model that the lottery cannot be evaluated far into the future\footnote{\new{In practice these are both desirable properties.
      The adversary should not learn if an honest party wins the lottery before that honest party has time to send out their block.
      Neither should the adversary be able to predict a sequence of slots that the they win.}}}.
We also do not model signatures.

Instead, we quantify our theorems in \cref{sec:results} by an appropriate hypothesis on the unforgeability of blocks produced by honest players (\cref{def:forging-free}).

\paragraph{Valid chains}
Our protocol has an initial block, $\GenesisBlock : \Block$, that all chains should end in and which we assume to have an honest baker identifier and the slot set to $0$. Using the lottery abstraction we define a valid chain.  
\begin{coqdef}{Valid chain}{BlockTree.v}{valid\_chain}{Protocol/BlockTree.v\#L51}
  We say that a chain is a valid chain if it fulfills the following three requirements
  \begin{itemize}
  \item All blocks in the chain need to be valid. A block $b$ is  valid if $\Winner\ (\bid\ b) (\slot\ b) = \top$. 
  \item The chain should be linked correctly: the field $\pred$ of a block contains a hash that is equal to that of the predecessor in the chain and the chain ends in the $\GenesisBlock$. 
  \item The projection of the fields $\slot$ from the chain forms a strictly decreasing sequence of slots. 
  \end{itemize}
\end{coqdef}
We define $\vc : \Chain \rightarrow \bool$, as a computable predicate ensuring these properties are fulfilled. 

\subsection{BlockTree}\label{sec:blockTree}
A NSB maintains a \emph{correct} tree of currently received blocks, from which the current \emph{best chain} can be derived.

\new{Previous analysis of NSB protocols~\cite{bbp15, bcuc17, our17, toychain18} provide an explicit algorithm for calculating the best chain from a set of chains and prove the security of this construction.}
Unfortunately, this approach creates a gap between the security analysis on an easily verifiable algorithms and the highly optimized code that is running in typical implementations of such a protocol.
The main performance bottleneck of the extracted implementation of \toychain~\cite{toychainThesis} is their blocktree which runs in $O (n^{4})$-time.
Comparable non-verified implementations run in $\sim O(n)$-time when the cost is amortized.

In this work, we take a different approach and specify the minimal requirements of a correct blocktree rather than providing an explicit construction for this data-structure.
\new{Taking this approach, we do not prove correctness of an efficient implementation.
  This could be done in two ways: 1) Either by providing a reference implementation (as in previous work) which can then be refined, or 2) by instantiating our abstract interface.
  We consider the second approach to be more flexible as it provides a \emph{minimal} specification.  
  Moreover, we do not explicitly provide an implementation of our specification. 
  We come back to this after \cref{def:self-contained}.\\

}

Following the style of \mathcomp we define a type, $\treeType$, that denotes a type that satisfies the requirements to achieve our security in the protocol.

\paragraph{Correctness conditions for a block tree}
For a type $\tv : \Type $ to be a $\treeType$, we demand that the following functions should be defined.

\new{
\begin{align*}
  &\treeI : \tv \\
  &\extendTree : \tv \rightarrow \Block \rightarrow \tv \\
  &\allBlocks : \tv \rightarrow \seq\ \Block \\
  &\bestChain : \Slot \rightarrow \tv \rightarrow \Chain
\end{align*}}
The function $\treeI$ corresponds to the requirement that there is an initial tree that the protocol can be instantiated with, $\extendTree$ gives a way to extend any tree returning a new tree, $\allBlocks$ should give a set of blocks that the tree has been extended with and finally $\bestChain$ allows one to extract what is currently the best chain of the tree with respect to a slot.

$T: \Type$ is a $\treeType$ if is instantiated, extendable, valid, optimal and self-contained:

\begin{coqdef}{Instantiated}{BlockTree.v}{all\_tree0}{Protocol/BlockTree.v\#L106}
  A type $T$ is \emph{instantiated} if no blocks are recorded in the initial structure except for $\GenesisBlock$ i.e.,
  \begin{equation*}
      \allBlocks\ \treeI \memeq \left[:: \GenesisBlock \right]. \label{c:emptytree}
  \end{equation*}
\end{coqdef}

\begin{coqdef}{Extendable}{BlockTree.v}{all\_extend}{Protocol/BlockTree.v\#L109}
  A type $T$ is \emph{extendable} if extending the structure with a block is recorded properly in the set of contained blocks i.e.,
  \begin{align*}
    \subAlign \forall (t : \tv) (b : \Block), \subBreak
    \subAlign \allBlocks (\extendTree\ t\ b) \memeq \allBlocks\ t \cat \left[:: b\right]. \label{c:all_extend}
  \end{align*}
\end{coqdef}

\begin{coqdef}{Valid}{BlockTree.v}{best\_chain\_valid}{Protocol/BlockTree.v\#L112}
  A type $T$ is \emph{valid} if the best chain achieved from this structure is always a valid chain i.e.,
  \begin{equation*}
    \forall (t : \tv) (sl : \Slot), \vc\ (\bestChain\ sl\ t). \label{c:best_chain_valid}
  \end{equation*}
\end{coqdef}

\begin{coqdef}{Optimal}{BlockTree.v}{best\_chain\_best}{Protocol/BlockTree.v\#L115}
  A type $T$ is \emph{optimal} if the best chain less than a slot achieved from this structure is at least as good as any other chain obtained from the set of blocks recorded in the structure i.e.,
  \begin{align*}
    \forall & (c : \Chain) (t : \tv) (sl : \Slot),  \\
            &\vc\ c \rightarrow \subBreak
            \subAlign c \subseteq \{b \in \allBlocks\ t \mid \slot\ b \leq sl\} \rightarrow \subBreak
            \subAlign |c|\leq |\bestChain\ sl\ t|. 
  \end{align*}
\end{coqdef}

\begin{coqdef}{Self-contained}{BlockTree.v}{best\_chain\_in\_all}{Protocol/BlockTree.v\#L118}\label{def:self-contained}
  A type $T$ is \emph{self-contained} if the best chain less than a slot achieved from this structure is a subset of the recorded blocks in the structure i.e.,
  \begin{align*}
    \subAlign \forall (t : \tv) (sl : \Slot), \subBreak
    \subAlign \bestChain\ sl\ t \subseteq \{ b \leftarrow \allBlocks\ t \mid \slot\ b \leq sl \}. \label{c:best_chain_in_all}
  \end{align*}
\end{coqdef}

\new{Note that a simple algorithm that keeps track of all possible chains that can be created from the received blocks and prunes these for blocks from future slots before calculating the best chain provides all of the desired properties. This algorithm is what is used in~\cite{op18}.}
  
Our development is parameterized over a specific implementation of such a type, $\Tree : \treeType$ that we use to build a particular tree, consisting of all blocks honest parties have received.

\subsection{Parties}\label{sec:parties}
We represent the knowledge of a participating party as a record containing their identity, a $\treeType$, and a blocktree of that type:
\begin{equation*}
  \LocalState \coloneqq \{ \id : \Party,\ \tT : \treeType,\ \tree : \tTT \}. 
\end{equation*}
We further parameterize our development by a tree implementation for each party, $\TreeTypeMap : \Party \rightarrow \treeType$.
Unlike traditional pen-and-paper proofs (and previous formalizations) this implies that our results in \cref{sec:results} are quantified over all parties using different implementations of the core data-structure.
This is a realistic scenario for a blockchain protocol, as parties might participate in the protocol with different devices and as a consequence different implementations optimized for their particular device.

Being able to make this quantification is another benefit of our abstract characterization of the core data-structure for the protocol.

\paragraph{Honest behavior}
The behavior of an honest party is defined by two stateful functions: One that defines an honest party's reaction when receiving a sequence of messages in a slot, \honestRcv, and one that defines what an honest party should do when baking for a slot, $\honestBake$.
Both functions take an argument of type $\LocalState$ and return an updated state together with a sequence of messages \new{(the type $\Messages$, see~\cref{sec:network})} that the party wishes to flood to other parties.

\new{\begin{align*}
  \honestRcv :\ & \Messages \rightarrow \subBreak
               \subAlign \Slot \rightarrow \subBreak
               \subAlign \LocalState \rightarrow \subBreak
               \subAlign (\Unit * \LocalState) \\
       \honestBake :\ & \Slot \rightarrow \subBreak
                        \subAlign \Transactions \rightarrow \subBreak
               \subAlign \LocalState \rightarrow \subBreak
               \subAlign (\Messages * \LocalState)
\end{align*}}
The honest parties receives in a straightforward manner, as they will simply extend their blocktree with all blocks they receive, using the $\extendTree$-function defined for their blocktree implementation.
When an honest party is invoked to bake they will test if they are the $\Winner$ of the current slot. 
\new{If so, they will calculate the best chain from their current block tree, disregarding blocks from future slots, and create a new block with the predecessor set to the hash of the head of the best chain. Then they will include the transactions provided as an argument in this block.}
Finally, they will extend their blocktree with this new block and create a message containing this block and flood this.\\

The honest behavior is computable, and to run the protocol these two functions could be extracted and connected to a network-shim\footnote{Code that floods messages as well as receives messages from other parties and invoking the $\honestRcv$.} and a time-shim\footnote{Code that invokes the $\honestBake$ each time a new slot start.}, similarly to what has been done for previous formalizations~\cite{kaizen19, toychainThesis}.

\paragraph{Adversarial parties}
We explicitly model an adversary within the system, by parameterizing the development by a type, \AdversarialState that the adversary can choose freely.
We furthermore let the adversary choose the behavior of any corrupted party by again parameterizing our development over two functions corresponding to the adversarial behavior when receiving blocks and when baking for a slot.
\begin{align*}
  \AdversarialRcv, \ifsubmission \ \ \ \ \else\fi \subAlign\subBreak
  \AdversarialBake :\ & \Slot \rightarrow\\
  & \Messages \rightarrow\\
  & \MsgTuples \rightarrow\\
  & \AdversarialState \rightarrow\\
  & (\seq\ (\Message * \DelayMap)\subBreak\subAlign * \AdversarialState) 
\end{align*}
\new{ The adversary's functions take more arguments than the corresponding honest ones.\footnote{\new{The adversary is not provided with any transactions as it can freely decide what to include in the blocks. Moreover, later we will quantify over any selection of transactions to honest parties (including over selection-algorithms that may be known to the adversary before hand).}}
  In this way we model a more powerful adversary by providing him with a complete view of the state: the entire history of messages sent in the system and those that are sent, but not yet delivered, as well as their delivery times (encapsulated in the type $\MsgTuples$; see~\cref{sec:network}).
  This type of powerful adversary, i.e.\ one that who has access to all messages sent even before they are delivered, is called a \emph{rushing} adversary.
  We also allow the adversary to supply an additional argument (of type $\DelayMap$) to the messages he wishes to be sent. This allows him a more fine-grained control over when his messages will be delivered (again see~\cref{sec:network}).
  Although the type-signatures of \AdversarialRcv and \AdversarialBake are similar, we parameterize our development by two distinct functions to make adversary much powerful as possible.
}

\new{Modelling an active adversary by quantifying over an opaque function was previously done in other \coq developments~\cite{fcf15,probchain18}.}

\subsection{Global state}
\label{sec:global-state}
We define a record type $\GlobalState$ that contains all the information for this protocol when it is executed. The $\GlobalState$ record has the following fields.

\begin{description}[leftmargin = \parindent, labelindent = \parindent, itemsep=1pt]
\item[Clock:] The current slot of the system. 
\item[Message buffer:] A buffer containing all messages that have been sent but not yet delivered in the system.
\item[State map:] A partial map of type $\Party \rightarrow \option\ \LocalState$ that keeps track of the local state of all participating parties.
\item[History:] The history of all messages that have been sent.
  This is merely a book-keeping tool for describing assumptions such as the absence of hash-collisions in the state.
  Examples of how this is used can be found in~\cref{sec:results}.
\item[Adversarial state:] The adversaries state.
\item[Execution order:] The order in which the system should activate its parties.
  This is merely a bookkeeping allowing the environment to decide the order of activations (see \cref{sec:reachable-worlds}).
\item[Progress: ] The progress that the system has made within a single slot
\begin{align*}
  \Progress \triangleq \{\Ready, \Delivered, \Baked\}. 
\end{align*}
How a global state can change its progress is defined in~\cref{sec:reachable-worlds}.
\end{description}

\subsection{Network}
\label{sec:network}
We assume a lock-step-synchronous network with a known upper bound on the delivery time. This is similar to what the first analysis of both PoW~\cite{bbp15} and PoS~\cite{our17} assumes. This can be extended to a semi-bounded delay network (with a known upper bound) in the same way as~\cite{bbp15,op18}. 
This network model is different from the analysis in~\cite{toychain18}, which assumed only a partially synchronous network\footnote{A network that only guarantees that messages eventually will be delivered.}. However, NSBs are not secure in that model.

More precisely, we assume that time is discretized into slots which are coarse enough for honest parties to have enough time to first execute their computations for a slot and then send out messages.
At this time there should be enough time left in the round such that any message sent out at this point is ready for the delivery phase of the next round.
This assumption enables the possibility of creating a flooding network with the property that if a message is sent by an honest party in slot $sl$ then it will be delivered to any other party at time $sl +1$.

Adversarial parties sending messages in slot $sl$ does, however, have the possibility of postponing sending their messages until the very end of the round in which case they can choose to let some honest parties receive their message in slot $sl+1$ and others in slot $sl+2$. \\

\new{At first this may seem as a stronger assumption than used in previous work~\cite{bbp15,our17,op18}. There adversaries can send different messages to different parties. Adversarial blocks will then be propagated to other honest parties only after an honest party extends these. This is because honest parties will send entire chains around instead of just blocks. Note, however, that our network model can easily be derived from their assumptions by simply letting all honest parties gossip about the blocks they receive.
  Our network model can be instantiated with a gossip protocol.
  This is closer to what is used in NSBs running in practice and more realistic than previous pen-and-paper modeling.\\
}

To capture this network in our formalization, we introduce the type \Message as an inductive type with only a single 
constructor namely $\BlockMsg : \Block \rightarrow \Message$, and the record \MsgTuple defined by
\begin{align*}
  \subAlign \MsgTuple \coloneqq \subBreak
  \subAlign \{ \msg : \Message,\ \rcv : \Party,\ \cd : \Delay \},
\end{align*}
where $\Delay \triangleq \{1,2\}$. The field $\msg$ contains the actual message that is to be delivered at the receiving party contained in the field $\rcv$.  
\cd is the current delay of the message, which will be decremented for all messages as time progresses in the model. 

The flooding network available to the parties is formalized as a set of functions that operate on a global state. The functionalities $\floodMsgs$ and $\floodMsgsAdv$ enable the honest parties, the adversary, respectively to send messages.
\begin{align*}
  \floodMsgs  :\ & \Messages \rightarrow \subBreak
                \subAlign \GlobalState \rightarrow \subBreak
                \subAlign \GlobalState\\
  \floodMsgsAdv :\ & \seq \ (\Messages * \DelayMap) \rightarrow \subBreak
                    \subAlign \GlobalState \rightarrow \subBreak
                \subAlign \GlobalState
\end{align*}
Both functions will create a new message-tuple with the message for each party in the execution order of the global state.
\floodMsgs  will set the delay of the messages that are being sent to $1$, whereas the \floodMsgsAdv takes an extra parameter for each message namely a $\DelayMap \triangleq \Party \rightarrow \{1,2\}$, such that the adversary for each message explicitly can choose what parties should have it delivered in the next round and what parties should have it delivered in two rounds.

\subsection{Reachable Worlds}
\label{sec:reachable-worlds}

To be able to reason about the reachable states of the protocol, we first define an initial global state, $\NI : \GlobalState$.
To this end we parameterize our development over a sequence of parties participating in the protocol, $\InitParties : \seq\ \Party$, and create an initial state for all these parties with their tree set to $\treeI$. The development is also parameterized over any initial state that an adversary wants to choose, $\AdvStateI : \AdversarialState$.

$\NI$ is now defined in a straightforward manner with no messages in the message-buffer, nothing in the history, $\AdvStateI$, and the parties' respective initial states.

We also parameterize our \new{development by a} total map $\Honest : \Party \rightarrow \bool$ which decides what function should be invoked for each respective party.
This corresponds to the adversary being able to statically decide who should be corrupted.

\new{We furthermore parameterize our development by a total map $\TxSelection : \Slot \rightarrow \Party \rightarrow \Transactions$ which decides what transactions honest parties should include in the blocks they bake.
  We choose this modelling as it is completely irrelevant for the blockchain \emph{what} payload parties make it carry.
  The entire proof could be (and was in earlier versions) performed without any content in the blocks.
  By adding \emph{some} payload inside blocks we allow the adversary the possibility to try to disturb the blockchain by letting (otherwise identical) blocks have different content.\footnote{\new{We are grateful to the CSF reviewers for this insight.}}\\}

To capture how the protocol progresses we define a relation over atomic steps of a global state that enforces a state-transition system. A depiction of the transition system can be found in \cref{fig:schedule}. In the definition below \emph{progress} refers to the progress stored in a global state.  

\begin{coqdef}{Atomic step reachable}{Schedule.v}{SingleStep}{Model/Schedule.v\#L98}
  \label{def:atomic-step-reachable}
  For any two states $\worldv_{1}, \worldv_{2} : \GlobalState$, we say that $\worldv_{2}$ is reachable in an atomic step from $\worldv_{1}$ if one of the following steps are taken.
  \begin{description}[leftmargin = \parindent, labelindent = \parindent, itemsep = 1pt]
  \item[Receive:] If the progress of $\worldv_{1}$ is \Ready, then $\worldv_{1}$ can step to the state obtained by invoking each respective parties delivery-function, update the state of the state according to the outcome of this, and set the progress to $\Delivered$. 
  \item[Bake:] If the progress of $\worldv_{1}$ is \Delivered, then $\worldv_{1}$ can step to the state obtained by invoking each \new{(honest or dishonest)} party's bake-function, updating the state according to this outcome, and setting the progress to $\Baked$. 
  \item[Increment:] If the progress of $\worldv_{1}$ is \Baked, then $\worldv_{1}$ can step to the state obtained by incrementing the slot number and updating the progress to $\Ready$.
  \item[Permute execution order:] Any $\worldv_{1}$ can step to the state obtained by permuting the execution order of $\worldv_{1}$. 
  \item[Permute message buffer:] Any $\worldv_{1}$ can step to the state obtained by permuting the message buffer of $\worldv_{1}$. 
  \end{description}
  When $\worldv_{1}$ can step to $\worldv_{2}$ in one atomic step, we write $\worldv_{1} \leadsto  \worldv_{2}$. 
\end{coqdef}
This transition relation can be seen as an environment activating the parties in a restricted order. We model a adversarial environment by allowing permutations of the message buffer and the execution order. This models a very powerful adversary who gets to choose the exact message order for all messages sent, and decides the execution order for each step\footnote{This is also our reason for representing the execution order and message buffer as lists rather than multisets as we wish to give the adversary as much power as possible, by letting him determine the exact order.}.
\cref{def:atomic-step-reachable} is formalized as an inductive relation over global states in \coq. \\

We extend this definition to cover multiple steps as the reflexive transitive closure of atomic steps. 
\begin{coqdef}{Reachable}{Schedule.v}{BigStep}{Model/Schedule.v\#L115}
  For any two states $\worldv_{1}, \worldv_{2} : \GlobalState$ we say that $\worldv_{2}$ is \emph{reachable} from $\worldv_{1}$ if $\worldv_{2}$ is reachable in zero or more atomic step from $N_{1}$.
  We write $\worldv_{1} \Downarrow \worldv_{2}$.
\end{coqdef}

\ifsubmission
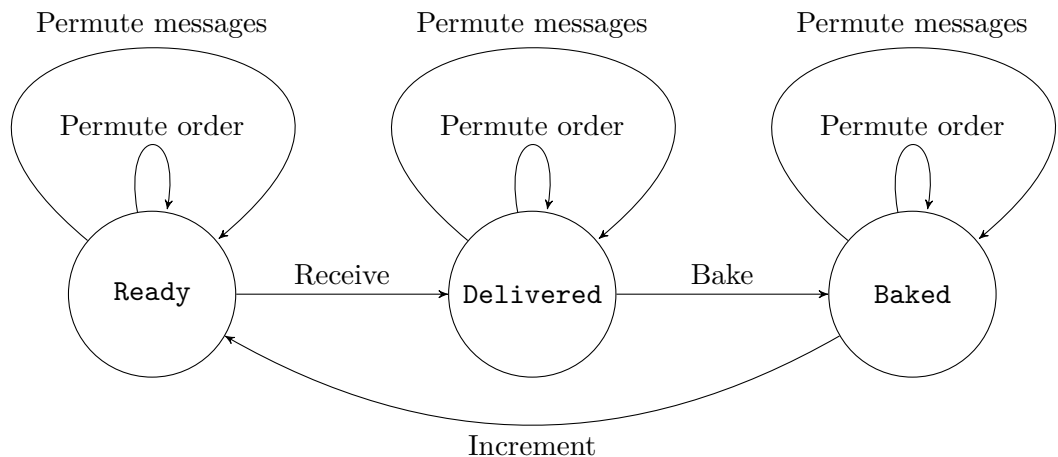
\begin{figure}[!htb]
  \centering
  \makebox[\columnwidth][c]{
    \begin{tikzpicture}
      [  ->
      , scale = 0.82
      , >=stealth'
      , outer loop/.style={out=130, in = 50, loop}
      , inner loop/.style={out=100, in = 80, loop}
      , transform shape
      ]
      \def\d{2}
      \node[circle, draw, minimum size = \d cm] at (-4,0) (R) {\Ready};
      \node[circle, draw, minimum size = \d cm] at (0,0) (D) {\Delivered};
      \node[circle, draw, minimum size = \d cm] at (4,0) (B) {\Baked};
      \draw (R) edge[above] node{Receive} (D)
      (D) edge[above] node{Bake} (B)
      (B) edge[bend left, below] node{Increment} (R)
      (R) edge[inner loop, above] node{Permute order} (R)
      (D) edge[inner loop, above] node{Permute order} (D)
      (B) edge[inner loop, above] node{Permute order} (B)
      (R) edge[outer loop, above] node{Permute messages} (R)
      (D) edge[outer loop, above] node{Permute messages} (D)
      (B) edge[outer loop, above] node{Permute messages} (B);
    \end{tikzpicture}
  }
  \caption{A depiction of the transition system that defines reachable states.}
  \label{fig:schedule}
\end{figure}
\else
\begin{figure}[!htb]
  \centering
  \makebox[\textwidth][c]{
    \begin{tikzpicture}
      [  ->
      , scale = 1
      , >=stealth'
      , outer loop/.style={out=140, in = 40, loop}
      , inner loop/.style={out=100, in = 80, loop}
      , transform shape
      ]
      \def\d{2.2}
      \node[circle, draw, minimum size = \d cm] at (-5,0) (R) {\Ready};
      \node[circle, draw, minimum size = \d cm] at (0,0) (D) {\Delivered};
      \node[circle, draw, minimum size = \d cm] at (5,0) (B) {\Baked};
      \draw (R) edge[above] node{Receive} (D)
      (D) edge[above] node{Bake} (B)
      (B) edge[bend left, below] node{Increment} (R)
      (R) edge[inner loop, above] node{Permute order} (R)
      (D) edge[inner loop, above] node{Permute order} (D)
      (B) edge[inner loop, above] node{Permute order} (B)
      (R) edge[outer loop, above] node{Permute messages} (R)
      (D) edge[outer loop, above] node{Permute messages} (D)
      (B) edge[outer loop, above] node{Permute messages} (B);
    \end{tikzpicture}
  }
  \caption{A depiction of the transition system that defines reachable states.}
  \label{fig:schedule}
\end{figure}
\fi
Our definition of reachable enforces that the set of parties participating in the protocol remains static through the execution of the protocol.

\section{Safety and Liveness}
\label{sec:results}
\new{This section will discuss our three main theorems (chain growth, chain quality and common prefix)} and outline the structure of their proofs.
The entire proof amounts to roughly $6$k lines of code using \mathcomp's compact proof language.

Throughout the section we make two standard assumptions about the transition system.
We assume that the list of parties participating ($\InitParties$) in the protocol is unique, i.e., that no party will be activated twice during the same atomic step, and that there is at least one honest party among the participants\footnote{This is not a requirement on the stake of the honest parties, but simply a requirement that at least one of the actual parties in the protocol behaves honestly.
  The requirements on the lottery and thus on the stake will appear as preconditions for the individual statements.
}.

\paragraph{Phrasing of theorems}
Our chain growth, chain quality and common prefix are stated as implications rather than the absolute probabilistic statements given in previous analysis.
Chain growth relies only on a certain number of lucky slots within the time-span of states, whereas chain quality relies on a collision-free state, a forging-free state and certain condition on the winning events in a time-span. Common prefix relies both on a collision-free and a forging-free state. It states that either the property holds or a bad event happens --- namely that the adversary has gotten an advantage that is statistically unachievable for a large $k$. 

\new{A probabilistic statement can be obtained by bounding the probabilities of the desired hypotheses (or conclusion).
  Formalizing this depends on the specific lottery functionality, the hash function, and the signature-scheme.
  This is not treated in this work, but below we will provide intuition how to prove this; see also \cref{sec:concrete-probs}.}
\subsection{Defining Preconditions}
\label{sec:preconditions}
We start by defining some basic concepts. First, we specialize hash-collisions to our setting.
Next, we state an assumption on the adversary's capability to publish blocks with honest identifiers, before we move on to define certain good and bad events with respect to the lottery.\\

Any NSB protocol only provides its guarantees under the assumption that there are no hash-collisions throughout the execution. We define this as a \emph{collision-free} state. 

\begin{coqdef}{Collision-free}{CQ.v}{collision\_free}{Properties/CQ.v\#L628}
 A global state $\worldv : \GlobalState$ with block history $bh : \seq\ \Block$ is \emph{collision-free} if
  \begin{align*}
    \subAlign \forall b, b' : \Block, \subBreak
    \subAlign b, b' \in bh \rightarrow \HashBlock\ b = \HashBlock\ b' \rightarrow b = b'.
  \end{align*}
\end{coqdef}
For any two global states $\worldv_{1}, \worldv_{2} : \GlobalState$, if $\worldv_{1} \Downarrow \worldv_{2}$ and $\worldv_{2}$ is collision-free, then $\worldv_{1}$ is also collision-free, as block histories are monotonously growing over reachable states.
We have taken care to phrase each of our main theorems using this definition, instead of assuming a global axiom on the injectivity of the hash-function or that any reachable state is collision-free.
\new{The introduction of such global axioms could lead to an inconsistency.
  Moreover, it would not be possible to bound the probability that such an axiom is satisfied by a collision-resistant hash-function.}
\new{We provide intuition how this can be done with the current formulation:
}

\new{
  \begin{remark}\label{rem:collision-free-game-relation}
    If $\worldv : \GlobalState$ is not collision-free, then two blocks were produced between the initial state $\NI$ and $\worldv$ where the hash-function collided.
    If an adversary can break the collision-free assumption with non-negligible probability, then one can construct a new adversary emulating both honest and dishonest players whom will produce a collision on the hash-function with non-negligible probability.
  \end{remark}
}

Another assumption that is needed in order to be able to state our main theorems is that the adversary cannot forge any honest blocks through the execution that led to a global state. We do not model signatures explicitly, so instead, we assume that the adversary cannot send out any block with the $\bid$-field set to the identifier of an honest party that is not already a part of the block history. 
\begin{coqdef}{Forging-free}{CQ.v}{forging\_free}{Properties/CQ.v\#L103}
  \label{def:forging-free}
  We say that a global state $\worldv : \GlobalState$ is \emph{forging-free} if for any activation of the adversarial functions, $\AdversarialBake, \AdversarialRcv$ with parameters from a global state $\worldv' : \GlobalState$ where $\worldv' \Downarrow \worldv$ implies that there are no honest blocks in what the adversary sends that is not already in the block history of $\worldv'$. 
\end{coqdef}
In order to state this in the formalization, we introduce a more fine-grained refinement of the reachable transition-relation. We need this to be able to precisely state that the assumption holds in between each individual party-activation and not only in the synchronous steps.
\\
The definition of forging-free closely corresponds to the property one could achieve by using an EUF-CMA (existential unforgeability under chosen message attack) secure signature scheme to sign blocks.
\new{
  \begin{remark}\label{rem:forging-free-euf-cma-relation}
    If $\worldv : \GlobalState$ is not forging-free, the adversary has been able to forge a message between the initial state $\NI$ and $\worldv$, and has thus succeeded in breaking the signature scheme.
    Any adversary that can break this assumption with a non-negligible probability will thus be able to break the EUF-CMA secure scheme with a non-negligible probability.    
  \end{remark}
}

We define a lucky slot to be any slot where an honest party wins the lottery and an adversarial slot to be the corresponding concept for adversarial parties. Finally, we define honest advantage to be the difference between these two amounts over a sequence of slots.

\begin{coqdef}{Lucky slot}{CG.v}{lucky\_slot}{Properties/CG.v\#L1354}\label{def:lucky-slot}
  A slot $sl$ is a \emph{lucky slot} if there is a party $p \in \InitParties$ s.t.\
  $\Winner\ p\ sl \land \Honest\ p$.
\end{coqdef}
\begin{coqdef}{Super slot}{CP.v}{super\_slot}{Properties/CP.v\#L41}\label{def:super-slot}
  A slot $sl$ is a \emph{super slot} if there is a exactly one party $p \in \InitParties$ s.t.\
  $\Winner\ p\ sl \land \Honest\ p$.
\end{coqdef}
\begin{coqdef}{Adversarial slot}{CQ.v}{adv\_slot}{Properties/CQ.v\#L124}\label{def:adv-slot}
  A slot $sl$ is an \emph{adversarial slot} if there is a party $p \in \InitParties$ s.t.\
  $\Winner\ p\ sl \land \neg \Honest\ p$.
\end{coqdef}
There is a close connection between a lucky slot and the creation of a \emph{left-isolated block} in the analysis of PoW~\cite{weight20}, as we have scaled our slots such that all honest blocks have time to propagate before the round begins. Similarly, super slots corresponds to \emph{isolated} blocks. We call the block won by an honest player in a super slot a \emph{super block}

\begin{coqdef}{Honest advantage}{CQ.v}{honest\_advantage\_range}{Properties/CQ.v\#L140}
  We define the \emph{honest advantage} for an interval of slots to be the difference between the number of lucky slots and the  number of adversarial slots in this period.
\end{coqdef}

\subsection{Preliminary Lemmas}
\label{sec:preliminary}
We now state some selected definitions and lemmas that are used to prove our main theorems. The first lemma we introduce describes how knowledge propagates between honest parties. 

\begin{coqlem}{Knowledge propagation}{CG.v}{honest\_tree\_subset}{Properties/CG.v\#L1324}
  \label{lem:knowledge-prop}
  Let $\worldv_{1}, \worldv_{2} : \GlobalState$ and $p_{1}, p_{2} : \Party$.
  If $\NI \Downarrow \worldv_{1}$, $\worldv_{1} \Downarrow \worldv_{2}$, $p_{1}$ is a party in $\worldv_{1}$ with tree $t_{1}$, $p_{2}$ is a party in $\worldv_{2}$ with tree $t_{2}$, $\worldv_{1}$ is at $\Ready$, $\worldv_{2}$ is at $\Delivered$, and $\worldv_{1}$ and $\worldv_{2}$ are in the same slot then
  \begin{equation*}
    \allBlocks\ t_{1} \subseteq \allBlocks\ t_{2}.
  \end{equation*}
\end{coqlem}
\begin{proof}[Proof sketch]
  Our main observation is that at any point in time a block is in the tree of $p_{1}$, it is either also already in $p_{2}$'s tree or to be delivered at the next delivery transition.
  Blocks can be added when an honest party wins the right to bake a block, in which case they will immediately send the block to all other parties and thus fulfill the invariant, or they can be added by an adversary and thereby delivered to an honest party by a delivery event, in which case it will be delivered to all other honest parties in the following delivery slot (by our network assumption).
  
  This is in particular true when $p_{1}$ and $p_{2}$ is at $\Ready$, which means that after the delivery transition $p_{2}$ will know all the blocks that $p_{1}$ knew before.
\end{proof}
Since honest parties extend their trees monotonously this subset-relation will also extend to any state that leads to $\worldv_{1}$ and any state that is reachable from $\worldv_{2}$.\\

The core insight of the proof for common prefix is that each time a super-slot is won the block produced in this slot will not have the same depth in a chain as any other honest block.
In order to define this precisely, we define how to calculate a chain from a block\footnote{This definition does not appear in previous pen and paper proofs, which only talks about positions of blocks without defining with respect to what set of blocks. }.

\begin{coqdef}{Chain from a block}{CP.v}{cfb}{Properties/CP.v\#L328}
  We define the chain from a block $b : \Block$ with respect to a sequence of blocks $bp : \seq\ \Block$ to be the chain obtained by following the pointers to from $b$ through $bp$ ending in $\GenesisBlock$.
  We write $\cfb\ b\ bp$ to denote this chain.
  If no such chain can be obtained by following pointers in $bp$ we say that $\cfb\ b\ bp = \emptyseq$.
\end{coqdef}
\begin{coqdef}{Position of a block}{CP.v}{pos}{Properties/CP.v\#L354}
  We furthermore define the \emph{position} of a block, written $\pos$, to be the length of the chain obtained by following the pointers from the block,
  \begin{equation*}
  \pos\ b \ bp \coloneqq |\cfb\ b \ bp|.
  \end{equation*}
\end{coqdef}

As this is not a structurally recursive function we use the \equations plugin~\cite{eq19} in order to automatically get a strong induction principle. This allows us to prove the following lemma that is a central step towards proving the common prefix property. 
\begin{coqlem}{Super block positions}{CP.v}{no\_honest\_pos\_share\_sb}{Properties/CP.v\#L1145}
  \label{lem:sb-pos}
  Let $\worldv : \GlobalState$, $sb, b : \Block$ and let $bh : \seq\ \Block$ be the history of blocks in $\worldv$.
  Suppose $\NI \Downarrow \worldv$, $\worldv$ is forging-free and collision-free, $b,sb \in bh$, $b$ is honest and $sb$ is a super block then
  \begin{equation*}
    \pos\ sb\ bp \neq \pos\ b\ bp. 
  \end{equation*}
\end{coqlem}
\begin{proof}[Proof sketch]
  The proof proceeds by induction on the transition relation $\NI \Downarrow \worldv$. The base case is trivial as there are no blocks in the block history of $\NI$. In the induction case we distinguish between which transition was taken last. 
  \begin{description}[leftmargin = \parindent, labelindent = \parindent, itemsep = 1pt]
  \item[Receive:] Receiving messages does not change the subset of the block history that is honest. Moreover, a collision-free state guarantees that the positions of the honest blocks that are already in the block history do not change. 
  \item[Bake:] Let $sl$ be the slot of $\worldv$. We note that any honest block $b'\in bh$ must have a slot number that is less than or equal to that of the current state, and distinguish between these two cases.
    \begin{description}[leftmargin = \parindent, labelindent = \parindent, itemsep = 1pt]
    \item[$\slot\ sb < sl$]: Any honest party that bakes a new block in this step must have known about $sb$ (by~\cref{lem:knowledge-prop}) and are aware of a valid chain that is at least as long as the position of $sb$. We will therefore have for any new block $b$ that is baked in such a way that $\pos\ sb\ bh < \pos\ b\ bh$. 
    \item[$\slot\ sb = sl$]: There is exactly one honest party that bakes a block in this step. By \cref{lem:knowledge-prop} this party must know about all other honest blocks baked in previous rounds. We will therefore have that for any old honest block $b$ that $\pos\ b\ bh  < \pos\ sb\ bh$. 
    \end{description}
  \item[Increment/Permute orders:] These transitions do not change the block history. \qedhere
  \end{description}
\end{proof}
At last we define pruning and a prefix, as well as a minor lemma relating the notions in order to phrase and prove our common prefix theorem.
\begin{coqdef}{Pruning}{CP.v}{prune\_time}{Properties/CP.v\#L1919}
  Let $c : \Chain$ be a chain and let $sl :\Slot$ be a slot.
  We prune $c$ by $sl$ by removing all blocks that has a slot higher than $sl$,
  \begin{equation*}
    \prune\ sl\ c \triangleq \{b \leftarrow c \mid \slot \ b \leq sl \}. 
  \end{equation*}
\end{coqdef}
For a valid chain, pruning corresponds to simply removing blocks until the head of the chain is below a or equal to a certain slot. We finally define prefix\footnote{Technically this is a \emph{suffix} due to the orientation of our list structure, but to avoid confusion we use the word prefix to align with previous results.}. 

\begin{coqdef}{Chain prefix}{SsrFacts.v}{suffix}{Properties/SsrFacts.v\#L294}
  Let $c_{1}, c_{2} : \Chain$.
  We say that $c_{1}$ is a \emph{prefix} of $c_{2}$ if there exists a $c_{3} : \Chain$ such that $c_{3}\cat c_{1} = c_{2}$.
  We write $c_{1} \suffix c_{2}$. 
\end{coqdef}

\begin{coqlem}{Prune prefix transitivity}{CP.v}{prune\_suffix\_trans}{Properties/CP.v\#L2433}
  \label{lem:prune-suffix-trans}
  For any $sl : \Slot$ and $c_{1},c_{2},c_{3}: \Chain$ such that $\prune\ sl\ c_{1}  \suffix c_{2}$ and $\prune\ sl\ c_{2} \suffix c_{3}$, we have $\prune\ sl\ c_{1}  \suffix c_{3}$. 
\end{coqlem}

\subsection{Main Theorems}
\label{sec:main-theorems}
We are now ready to state our three main theorems. For clarity we ignore the constants $-1$ and $1$ when counting the number of lucky/adversarial/super slots.
These constants are used to account for adversary's ability to wait one more round to bake than the honest parties, because he immediately knows of all previously baked blocks.
\new{The precise statements can be found in the accompanying formalization. }\\

At a slot $sl$ any party with a tree $t$ will consider their best chain to be the chain calculated from the tree by disregarding all blocks from this slot and the future, $\bestChain\ (sl-1)\ t$. 
We will show the three key properties for such chains.\\

The chain growth property intuitively says that in each period, the best chain of any honest party will increase at least by a number that is proportional to the number of lucky slots in that period.

\begin{coqthm}{Chain Growth}{CG.v}{chain\_growth\_parties}{Properties/CG.v\#L1887}
  \label{thm:cg} 
  Let $\worldv_{1}, \worldv_{2} : \GlobalState$, $p_{1}, p_{2} : \Party$, $sl_{1}, sl_{2} : \Slot$ and $w : \nat$.
  If $\NI \Downarrow \worldv_{1}$, $\worldv_{1} \Downarrow \worldv_{2}$, $p_{1}$ is a party in $\worldv_{1}$ with tree $t_{1}$, $p_{2}$ is a party in $\worldv_{2}$ with tree $t_{2}$, the round of $\worldv_{1}$ is $sl_{1}$, the round of $\worldv_{2}$ is $sl_{2}$ and there are at least $w$ lucky slots between $\worldv_{1}$ and $\worldv_{2}$ then
  \begin{equation*}
    |\bestChain\ (sl_{1} -1)\ t_{1}| + w \leq |\bestChain\ (sl_{2} -1) \ t_{2}|.
  \end{equation*}
\end{coqthm}

\begin{proof}[Proof sketch]
  We proceed by induction on the number of lucky slots, $w$.

  The base case follows by monotone growth of honest chains over time\footnote{Technically, the slot number of $\worldv_{1}$ needs to be strictly smaller than that of $\worldv_{2}$, as the knowledge of $p_{1}$ needs to have time to propagate to $p_{2}$ by \cref{lem:knowledge-prop}.}.
  In the induction case we identify the global state $\worldv$ with the lowest slot number $sl$ s.t., $\worldv_{1}\Downarrow \worldv$, $\worldv \Downarrow \worldv_{2}$, and $\luckySlot\ sl$.
  In the global state $\worldv$, we establish that the honest party who wins the slot creates a new chain that is strictly longer than any chain of an honest party in $\worldv_{1}$, as they knew what was there before by~\cref{lem:knowledge-prop}. We complete the proof by applying the induction hypothesis to $\worldv$.
\end{proof}
\new{For a concrete lottery implementation, a probabilistic version of \cref{thm:cg} can be proved by calculating the expected number of lucky slots in a period and then using the Chernoff-bound to upper-bound the likelihood that less lucky slots than expected occur.\\}

We now present the chain quality property.
\new{The chain quality property says intuitively that within any chunk of consecutive blocks in an honest party's best chain, there is an honest share of blocks. This share is proportional to the difference between the number of honest and adversarial slots.}

\begin{coqthm}{Chain Quality}{CQ.v}{chain\_quality}{Properties/CQ.v\#L1224}
  \label{thm:cq}
  Let $\worldv : \GlobalState$, $p : \Party$ and $w : \nat$. Suppose $\NI \Downarrow \worldv$, $\worldv$ is forging-free and collision-free, $p$ is a party in $\worldv$ with tree $t$, the round of $\worldv$ is $sl$, and let $B_{i}\dots B_{j}$ be a consecutive interval of blocks of $\bestChain\ (sl -1) \ t$. If there is an \emph{honest advantage} of at least $w$ for time periods longer than $\slot\ B_{i} - \slot\ B_{j}$ then the number of honest blocks in $B_{i}\dots B_{j}$ will be at least $w$. 
\end{coqthm}
\begin{proof}[Proof sketch]
  We define $B_{\hat{i}}$ and $B_{\hat{j}}$ s.t.
  $B_{\hat{i}}\dots B_{\hat{j}}$ is the smallest interval of $\bestChain\ (sl -1) \ t$ such that $B_{i}\dots B_{j} \subseteq B_{\hat{i}}\dots B_{\hat{j}}$, $B_{\hat{i}}$ is honest and $B_{\hat{j}}$ is either honest or the head of $\bestChain\ (sl-1) \ t$ \footnote{$B_{\hat{i}}$ is well defined as we consider the genesis block to be honest.}.
  As $B_{\hat{i}}$ is honest, we can apply~\cref{thm:cg} to establish that $|B_{\hat{i}}\dots B_{\hat{j}}|$ is at least the number of adversarial slots in the time span between the creation of $B_{\hat{i}}$ and $B_{\hat{j}}$ plus the honest advantage in this time span.
  As all blocks in a valid chain (and as $\bestChain\ (sl-1) \ t$ is valid) have unique slot numbers this implies that the there must be at least $w$ honest blocks in between $B_{\hat{i}}$ and $B_{\hat{j}}$ and therefore also $w$ honest blocks in $B_{i}\dots B_{j}$.
\end{proof}
We achieve the full chain-quality property that is defined for any fragment of any honest party's best chain rather than the somewhat weaker property considered in~\cite{ren19}. 

\new{A probabilistic version of \cref{thm:cq} can be proved for a lottery where the expected number of lucky slots is higher than the expected number of adversarial slots.
  This induces the assumption that a majority of stake is to be honest.
  If this is the case, then a standard probability bounds (such as Chernoff's) can be used to bound the likelihood that less lucky, respectively more adversarial, slots occur than expected within a period of slots.}

Together chain growth and chain quality prove liveness, as chain growth ensures that more blocks will be appended to any honest party's log and chain quality ensures that there will be some honest input to this log. \\ %

The common prefix property informally says that during the execution of the protocol the chains of honest parties will always be a common prefix of each other (after removing some blocks on the chain).
We follow~\cite{weight20, bbp15} and define two variants of the common prefix property.
The first variant ensures that any two best chains of honest parties are consistent within a single round, and the second variant ensures that the best chain of an honest party is consistent with earlier best chains of any honest party. The latter variant constitutes safety for blockchain consensus protocols.
\begin{coqlem}{Common prefix-lemma}{CP.v}{\new{cp\_prune\_gen\_inc}}{Properties/CP.v\#L2343}
  \label{lem:cp-lem}
  Let $\worldv : \GlobalState$, $p : \Party$, $c : \Chain$, $k : \Slot$ and $bh : \seq\ \Block$.
  Suppose $\NI \Downarrow \worldv$, $\worldv$ is forging and collision-free, $p$ is a honest party in $\worldv$ with tree $t$, the round of $\worldv$ is $sl$, the block history of $\worldv$ is $bh$, $c \subseteq bh$, that $c$ is a valid chain, all blocks in $c$ have a slot number less than $sl$ and that $|\bestChain\ (sl-1) \ t| \leq |c|$.
  Then one of the following events occurs:
  \begin{enumerate}
  \item $\prune\ k\ (\bestChain\ (sl-1) \ t) \suffix c$ \label[event]{itm:cp-good}
  \item There exists $sl' : \Slot$, s.t.\ $sl' \leq k$ and the number of super slots in the slot range from $sl'$ to $sl$ is less than two times the number of adversarial slots in the same period of time. \label[event]{itm:cp-bad}
  \end{enumerate}
\end{coqlem}
\begin{proof}[Proof sketch]
  We define $b'$ to be first honest block in the common stem of $c$ and $\bestChain\ (sl-1) \ t$. If $k < \slot\ b'$ we can conclude $\prune\ k\ (\bestChain\ (sl-1) \ t) \suffix c$. Otherwise we show~\cref{itm:cp-bad}.

  We define $sl'$ as $\slot\ b'$. Let $bh$ be the block history of $\worldv$. 
  For any honest block $b$ that is produced between $\slot\ b'$ and $sl$, we have
  \begin{equation*}
    \pos\ b'\ bp < \pos\ b\ bp \leq |\bestChain\ (sl-1) \ t| \leq |c|.
  \end{equation*}
  $\pos\ b'\ bp < \pos\ b\ bp$ because at the time $b$ was created the honest party that created it knew about a chain of length $\pos\ b' \ bp$, and $\pos\ b\ bp \leq |\bestChain\ (sl-1) \ t|$ as otherwise there would be a longer chain available to $p$ at time $sl$.
  Any adversarial slot can appear at most once on each chain. So, by~\cref{lem:sb-pos} there must be an adversarial slot for every two super blocks in this period.
\end{proof}
  \begin{remark}\label{rem:cp-lem-parties}
    For any reachable $\worldv$ global state with two honest parties, \cref{lem:cp-lem} can be instantiated with $c$ being the longer of the best chains for these parties.
    This will thus give us that the best chain of any honest party will be a prefix of any other honest party's best chain.
  \end{remark}

  \begin{coqthm}{Timed Common prefix}{CP.v}{timed\_common\_prefix}{Properties/CP.v\#L2499}
    \label{thm:tcp}
  Let $\worldv_{1}, \worldv_{2} : \GlobalState$, $p_{1}, p_{2} : \Party$, $sl_{1}, sl_{2} : \Slot$ and $k : \Slot$.
  If $\NI \Downarrow \worldv_{1}$, $\worldv_{1} \Downarrow \worldv_{2}$, $\worldv_{2}$ is forging-free and collision-free, $p_{1}$ is a party in $\worldv_{1}$ with tree $t_{1}$, $p_{2}$ is a party in $\worldv_{2}$ with tree $t_{2}$, the round of $\worldv_{1}$ is $sl_{1}$ and the round of $\worldv_{2}$ is $sl_{2}$.
  Then one of the following events occurs:
  \begin{enumerate}
  \item $\prune\ k\ (\bestChain\ (sl_{1}-1) \ t_{1}) \suffix (\bestChain\ (sl_{2}-1) \ t_{2})$ \label[event]{itm:tcp-good}
  \item There exists $sl',sl'' : \Slot$, s.t. $sl' \leq k$, $sl_{1} \leq sl'' \leq sl_{2}$ and that the number of super slots in the slot range from $sl'$ to $sl''$ is less than two times the number of adversarial slots in the same period of time.  \label[event]{itm:tcp-bad}
  \end{enumerate}
\end{coqthm}
\begin{proof}[Proof sketch]
  The proof proceeds by induction on the transition relation $\worldv_{1}\Downarrow \worldv_{2}$. The base case where $\worldv_{1} = \worldv_{2}$ is solved by applying~\cref{lem:cp-lem} (in particular~\cref{rem:cp-lem-parties}). In the induction case we distinguish between which transition was taken last.
  \begin{description}[leftmargin = \parindent, labelindent = \parindent, itemsep = 1pt]
  \item[Receive:] The induction hypothesis gives us that the statement is true for the tree $t_{2}'$ which $p_{2}$ has just before he receives the messages in this round.
    The messages that $p_{2}$ receives in this round must however already be in the block history and therefore~\cref{lem:cp-lem} can be applied.
    This either results in~\cref{itm:tcp-bad} or we can apply~\cref{lem:prune-suffix-trans} to achieve that $\prune\ k\ (\bestChain\ (sl_{1}-1) \ t_{1}) \suffix (\bestChain\ (sl_{2}-1) \ t_{2})$.
  \item[Bake:] The induction hypothesis gives us that the statement is true for the tree $t_{2}'$ which $p_{2}$ has just before he tries to bake for this slot. If $p_{2}$ bakes a block for the slot $sl$, the new block that is baked cannot itself be a part of the $\bestChain\ (sl_{2}-1)\ t_{2}$ but it might however still change the internal structure of the $t_{2}$ such that $\bestChain\ (sl_{2}-1)\ t_{2} \neq \bestChain\ (sl_{2}-1)\ t_{2}'$. This new chain must, however, already be a part of the block history, and therefore~\cref{lem:cp-lem} can be applied.
    This either results in~\cref{itm:tcp-bad} or we can apply~\cref{lem:prune-suffix-trans} to achieve that $\prune\ k\ (\bestChain\ (sl_{1}-1) \ t_{1}) \suffix (\bestChain\ (sl_{2}-1) \ t_{2})$.
  \item[Increment:] Incrementing the time allows for a slightly longer best chain than just before time was incremented.
    We apply the induction hypothesis to establish the relationship between the old best chain of $p_{2}$ and the best chain of $p_{1}$.
    Now we again apply~\cref{lem:cp-lem} and~\cref{lem:prune-suffix-trans}.
  \item[Permute execution order/message buffer:] These transitions do not change the best chains of any honest parties and the induction hypothesis can be applied. \qedhere
  \end{description}
\end{proof}

\new{As the conclusion of \cref{thm:tcp} is a disjunction it is enough to exclude \cref{itm:tcp-bad} from happening to ensure \cref{itm:tcp-good}.
  To achieve a probabilistic bound for \cref{itm:tcp-bad}, it is necessary that the lottery ensures that the expected amount of super-slots is more than twice the expected amount of adversarial slots.\footnote{For this to be possible for a concrete lottery construction, such as the one in Ouroboros Praos, at least $\frac{2}{3}'s$ of the underlying stake needs to be controlled by honest parties.}
  If that is the case, standard probability bounds (such as Chernoff's) can again be used to upper-bound the likelihood that less super slots, respectively more adversarial slots, than expected occur within a period of slots.
  Finally, to exclude that \emph{any} such period exists, union-bound is used to sum the probabilities of all the different interval lengths larger than $k$ but less than the current slot number.
}

A \emph{covert adversary} is one that leaves no trace that it did not follow the protocol. Such adversary would only be able to place each block on one chain. If we restrict ourselves to such adversaries, we 
\new{would immediately obtain a tighter bound. We could follow~\cite{weight20} and only need to assume that a majority of the resources behaves honest.}

\section{Related Work}
\label{sec:related-work}

\paragraph{Verified distributed systems}
A series of works have focused on formally verifying distributed systems in a non-Byzantine setting.
Raft~\cite{raft14} is a consensus algorithm that withstands benign failures and is simpler than similar algorithms, such as Paxos.
The safety property of Raft was formalized using the Verdi framework~\cite{verdi15, verdi16}.
Verdi relies on a shallow-embedding of protocols into \coq and provides the verified-system-transformers which facilitate composable verification.
Applying \coq's extraction to the Raft consensus protocol one obtains an implementation when connected to a network-shim. Their extracted code is as efficient as non-verified implementations.

Disel~\cite{disel17} is a framework for verifying distributed systems. It is built on a foundation of separation logic embedded in \coq and allows verifying \ocaml like programs using a Hoare style reasoning.
One can use the partial correctness of their Hoare style specifications to reason about safety.
Aneris~\cite{aneris20} is another framework embedded in \coq for verifying distributed systems.
It is built upon the Iris separation logic~\cite{iris15}, which allows reasoning about multi-threaded computations for local nodes while being able to combine the statements about local nodes to safety statements for the entire system.
Neither Disel nor Aneris has been used to reason about Byzantine behavior.

Lamport designed TLA+~\cite{tlap93} with the specific purpose of formally specifying and checking distributed protocols. This was used together with the TLAPS model-checker to check the safety (but not liveness) of PBFT~\cite{tlabft11}. 

IronFleet~\cite{ironfleet15} is a framework for combining both TLA-style specifications that are machine-checkable and Hoare style specifications in Dafny~\cite{dafny10}.
They prove a performant implementation of Paxos (a consensus algorithm withstanding benign failures) to be both safe and live.

\paragraph{Verified cryptographic protocols}
\new{There is an impressive amount of work verifying cryptographic primitives and two-party protocols~\cite{sok19}. 
However, there are only few works that verify multi-party protocols that are designed to be robust in an adversarial setting.
We mention the formalizations of multiparty computation~\cite{mpcec} and the AWS key-server~\cite{AWS}. These are both done in \easycrypt in the computational model. }
These works benefits from \easycrypt's logic that allows to reason about game-hops easily but also show limitations of \easycrypt's build-in programming language \texttt{pwhile} that lacks primitives for communication. The latter increases the complexity of the formalizations.

Modern cryptographic security proofs of consensus, e.g.~\cite{bcuc17, og18}, emphasize the use of an informal composible framework.
This will also be important for us when we want to prove that the system remains secure when we instantiate our lottery functionality with an implementation that has been proved to be secure in isolation.
Fortunately, such modular/composible frameworks are being developed more formally~\cite{EasyUC,CCcryptHOL}. However, only very simple protocols have been proven secure using these, due to the complexity of the frameworks themselves. 

\section{Conclusion}
\label{sec:concl-future-work}
We have given a formalized proof that a PoS NSB protocol with a static set of corrupted parties in a synchronous network has chain growth, chain quality, and common prefix.
This has required us to define precise semantics for the execution of the protocol.
We have defined honest behavior by computable functions and used this to define a relation on reachable global states.
We have also developed a new methodology for specifying core data-structures by their functional behavior rather than concrete implementation.
\new{This enables us to focus on the core combinatorial arguments while also providing a clear specification for optimized implementation.}
The methodology further has the consequence that we are able to prove security for parties running different implementations of the same protocol.

\new{\paragraph{Acknowledgments}
We thank Ilya Sergey for helpful discussions on \toychain and its variants, Daniel Tschudi for discussions in the beginning of the project, Thomas Dinsdale-Young for providing helpful insights into the Concordium Blockchain implementation, Jesper Buus Nielsen for helping to clarify the cryptographic models and ideas to the simple proof of common prefix, and Sabine Oechsner for valuable discussions and feedback. 
}

\bibliographystyle{alpha}
\bibliography{vpos}

\appendix
\ifsubmission\else {\Huge Appendix} \fi

\new{
  \ifsubmission
  \subsection{Concrete Probability Bounds}\label{sec:concrete-probs}
  \else \section{Concrete Probability Bounds}\label{sec:concrete-probs}
  \fi
For parties running the blockchain quantitative guarantees will often be more useful than the implications stated in \crefrange{thm:cg}{thm:tcp}.
What is the minimal expected growth of the best chain?
How long does a party need to wait before it is $99\%$ certain that a block will not be rolled back?

To answer these questions, we will show how to bound the probabilities of the preconditions/conclusions of \crefrange{thm:cg}{thm:tcp}. We will not discuss the probability of having a \emph{forging-free} and \emph{collision-free} global state any further as we have already done so in \cref{rem:collision-free-game-relation} and \cref{rem:forging-free-euf-cma-relation}. Instead, we focus on the probability that a sequence slots occurs that fulfills the respective preconditions or excludes a part of the conclusion.
\\

\crefrange{thm:cg}{thm:tcp} hold for any abstract lottery function, thus in particular for a random \emph{function}\footnote{\new{I.e., a computation that, when evaluated throughout the execution of the protocol, returns the same output on same inputs.}}.
Hence, the properties also hold for an implementation of the lottery such as the one constructed in Ouroboros Praos~\cite{op18}.
The lottery in Ouroboros Praos relies on a VRF. This is where the probabilities arises.
For simplicity let us assume that a concrete lottery gives rise to a series of independent random variables (as the one from Ouroboros Praos) corresponding to whether a specific slot fulfills \crefrange{def:lucky-slot}{def:adv-slot},
\begin{align*}
  &\rvarLS_{i}=
  \begin{cases}
    1 \text{ if slot } i \text{ is a lucky slot} \\
    0\text{ else}
  \end{cases}\\
  &\rvarSS_{i}=
  \begin{cases}
    1 \text{ if slot } i \text{ is a super slot} \\
    0\text{ else}
  \end{cases}\\
  &\rvarAS_{i}=
  \begin{cases}
    1 \text{ if slot } i \text{ is a adversarial slot} \\
    0\text{ else}
  \end{cases}.
\end{align*}
Given that a lottery gives rise to such a random experiment, we now wish to bound the probability that a certain sequence of slots satisfies the preconditions/conclusions of \crefrange{thm:cq}{thm:tcp}. 
Before we proceed to bounding the probabilities for such a lottery construction, we record a standard probability bound.
\begin{lem}{Chernoff} \label{lem:Chernoff}
  Let $X_1, \ldots, X_n$ be independent random variables with $X_i \in \{0,1\}$ for all $i$, and let $\mu \coloneqq \Ex \bigl[\sum_{i=1}^{n} X_i \bigr]$. We then have for all $\delta \in [0,1]$,
  \begin{equation*}
    \Pr \Biggl[\sum_{i=1}^{n} X_i \leq (1 - \delta) \mu \Biggr] \leq e^{- \frac{\delta^2 \mu}{2}},
  \end{equation*}
  and
  \begin{equation*}
    \Pr \Biggl[\sum_{i=1}^{n} X_i \geq (1 + \delta) \mu \Biggr] \leq e^{- \frac{\delta^2 \mu}{3}}.
  \end{equation*}
\end{lem}
We also introduce convenient notation for the successes of the variables $\pLS \coloneqq \Pr[\rvarLS_{i} = 1]$,  $\pLS \coloneqq \Pr[\rvarSS_{i} = 1]$, and $\pAS \coloneqq \Pr[\rvarLS_{i} = 1]$. For an interval of slots $r$ we define
\begin{equation*}
  \rvarLS (r) = \sum_{i \in r} \rvarLS_{i}, \aquad \rvarSS (r) = \sum_{i \in r} \rvarSS_{i}, \aquad \text{and } \rvarAS (r) = \sum_{i \in r} \rvarAS_{i},
\end{equation*}
and the corresponding expected values
\begin{equation*}
  \Ex[\rvarLS (r)] = r\cdot \pLS, \aquad \Ex[\rvarSS (r)] = r \cdot \pSS, \aquad \text{and } \Ex[\rvarAS (r)] = r \cdot \pAS. 
\end{equation*}
By instantiating \cref{lem:Chernoff} for these specific variables, we now have that for all $\delta_{1}, \delta_{2}, \delta_{3} \in [0,1]$,
\begin{align}
  \Pr[\rvarLS (r) \leq (1- \delta_{1})\cdot r\cdot \pLS ] \leq e^{-\frac{\delta_{1}^{2}\cdot r \cdot \pLS}{2}}, \label{eq:ls-bound} \\
  \Pr[\rvarSS (r) \leq (1- \delta_{2})\cdot r\cdot \pSS ] \leq e^{-\frac{\delta_{2}^{2}\cdot r \cdot \pSS}{2}}, \label{eq:ss-bound}\\
  \Pr[\rvarAS (r) \geq (1 + \delta_{3})\cdot r\cdot \pAS] \leq e^{-\frac{\delta_{3}^{2}\cdot r \cdot \pAS}{3}}. \label{eq:as-bound}
\end{align}
Using these we now show how to bound the probabilities for chain growth and common prefix. 
\paragraph{Chain Growth}
\cref{eq:ls-bound} provides a lower bound on the number of lucky slots as a function of the interval length.
As \cref{thm:cg} ensures chain growth corresponding to this quantity, this provides a lower bound on the chain growth as a function of the interval length.

\paragraph{Common Prefix}
For common prefix we wish to exclude that \cref{itm:tcp-bad} from \cref{thm:tcp} happens.
To do so we need that
\begin{equation*}
  \rvarSS (r) > 2 \cdot \rvarAS (r). 
\end{equation*}
For all $\delta, \delta' \in [0,1]$, we have that $ \rvarSS (r) > (1- \delta)\cdot r\cdot \pSS$ except with probability $e^{-\frac{\delta^{2}\cdot r \cdot \pSS}{2}}$. Except with probability $e^{-\frac{\delta'^{2}\cdot r \cdot \pAS}{3}}$, we have that $(1 + \delta')\cdot r\cdot \pAS > \rvarAS (r)$.
So we need to ensure that
\begin{equation*}
  (1- \delta)\cdot r\cdot \pSS > 2 \cdot (1 + \delta')\cdot r\cdot \pAS.
\end{equation*}
To do so we need the assumption on the lottery that $\exists \epsilon, \pSS \geq 2 \cdot \pAS + \epsilon $\footnote{This corresponds to assuming that $\nicefrac{2}{3}$ of the stake behaves honestly for the Ouroboros Praos lottery.}.
This implies the following condition
\begin{align}
  &(1- \delta)\cdot r\cdot (2 \cdot \pAS + \epsilon) > 2 \cdot (1 + \delta')\cdot r\cdot \pAS \nonumber\\
  &\Updownarrow\nonumber\\
  &\epsilon > \left(\frac{(1 + \delta')}{(1- \delta)} - 1\right)\cdot \pAS\cdot 2\label{eq:common-cp}.
\end{align}
This can be satisfied by choosing $\delta$ and $\delta'$ to be small.
The probability that \cref{eq:common-cp} does not hold decrases exponentially with $r$. To be precise as
\begin{equation*}
  e^{-\frac{\delta^{2}\cdot r \cdot \pSS}{2}} + e^{-\frac{\delta'^{2}\cdot r \cdot \pAS}{3}}.
\end{equation*}
To bound the existence of a time interval larger than a specific $r$, but less than the current world length (and thus exclude \cref{itm:tcp-bad}), we use union-bound and take the sum of these exponentially decreasing probabilities. 

\paragraph{Chain Quality}
Can be proved by the exact same approach as the common prefix, by using lucky slots instead of super slots and assuming only an honest majority. 
}

\old{
\subsection{Formalizing probabilities}
  The properties holds in particular for lottery, $\mathtt{L}$ which would rely on a hash-function $h: \{0,1\}^{*} \rightarrow \{0,1\}^{k}$ and a signature scheme. More importantly, the properties would also hold when $\mathtt{L}$ is analyzed in the RO model (where $h$ is replaced by an actual RO).

Now analyzing $\texttt{L}$ in the RO model we could show that this construction satisfies the probabilities presented in the next section.

\subsubsection{Eager vs. Lazy Sampling}
The RO model works by letting all parties have access to a the same random oracle. There are two options to model a common random oracle like this:
\begin{description}
\item[Multiple-draws:] We let the RO be a state-full probabilistic computation of the type 
  \begin{align*}
    \RO : \nat * \ROState \rightarrow \mathbb{G}(\{0,1\}^{k}* \ROState).
  \end{align*}
  The idea is that each time the $\RO$ is queried with it checks the state if the query was already made before, and if it is it returns the same value as returned earlier. Otherwise it randomly draws a uniform bit string in $\{0,1\}^{k}$ returns this and an updated state.
\item[One-draw:] The other option is to let the RO be evaluated just once before the protocol is ran. I.e., draw a \emph{particular} RO from the distribution
  \begin{align*}
    \RO \leftarrow\!\!\!\!\$\ \mathbb{G} (\nat_{\leq s} \rightarrow \{0,1\}^{k}). 
  \end{align*}
  This is at simplest model with a finite domain.
\end{description}

In \cite{sha3} a distinction is made between a an eager and a lazily evaluated lottery which are shown to be indistinguishable. Both these models lets the lottery be have the type
\begin{align*}
  \RO : \nat * \ROState \rightarrow \mathbb{G}(\{0,1\}^{k}* \ROState),
\end{align*}
but the difference between a eager and a lazy RO in their model is that the eager one, accepts an additional \emph{sample(x)}-query which makes it sample a particular value $x$ ahead of time. Their model is however more complicated as we do not need the \emph{set} and \emph{remove} functions on the lottery\footnote{These two functions are what gives the notion of a programmable random oracle.}.\\

\paragraph{Combining with current formalization}
We opt for using a combination of the two types of lotteries, by defining a lottery that is eagerly evaluated on a finite prefix of time. In particular, we define a total lottery with the type:
\begin{align*}
  \TotalWinner :\ & (n : \nat) \rightarrow \\
                 & (L_{1} : \Party \rightarrow \Slot_{\leq n} \rightarrow \bool) \rightarrow \\
                  & (L_{2} : \Party \rightarrow \Slot_{> n} \rightarrow \bool) \rightarrow \\
                  & (\Party \rightarrow \Slot \rightarrow \bool) \coloneqq \\
                  & \lambda n.\ \lambda L_{1}.\ \lambda L_{2}. \lambda sl.\ \lambda p. \\
                  & \textbf{ if } n \leq sl \textbf{ then } L_{1}\ sl\ p\textbf{ else } L_{2}\ sl\ p
\end{align*}
The idea is that we will instantiate this lottery with a sufficiently large random draw such that the total lottery will be random for the finite prefix that determines the value of the lottery for the finite slots that we care for. For any $n : \nat$ we let $D_{n}$ be a distribution over lotteries for evaluable on the finite prefix up to $n$ slots. We also index our big step semantics by the lottery which it is ran and write $\Downarrow_{l}$ for the big step semantics using lottery $l : \Party \rightarrow \Slot \rightarrow \bool$. We are now able to state the probabilistic version of timed common prefix. 
  \begin{coqthm}{Probabilistic Timed Common prefix}{Not in Coq yet}{}{}
    Let $\worldv_{1}, \worldv_{2} : \GlobalState$, $p_{1}, p_{2} : \Party$, $sl_{1}, sl_{2} : \Slot$ and $k : \Slot$, where $p_{1}$ is a party in $\worldv_{1}$ with tree $t_{1}$, $p_{2}$ is a party in $\worldv_{2}$ with tree $t_{2}$, the round of $\worldv_{1}$ is $sl_{1}$ and the round of $\worldv_{2}$ is $sl_{2}$.
    If $L_{1}$ is drawn from $D_{sl_{2}}$ then for all $L_{2} : \Party \rightarrow \Slot_{>sl_{2}} \rightarrow \bool $ then if $\NI \Downarrow_{(\TotalWinner\ L_{1}\ L_{2})} \worldv_{1}$, $\worldv_{1} \Downarrow_{(\TotalWinner\ L_{1}\ L_{2})} \worldv_{2}$, and $\worldv_{2}$ is forging-free and collision-free, 
    then 
    $$\prune\ k\ (\bestChain\ (sl_{1}-1) \ t_{1}) \suffix (\bestChain\ (sl_{2}-1) \ t_{2})$$
    unless with probability equal to that there exists $sl',sl'' : \Slot$, s.t. $sl' \leq k$, $sl_{1} \leq sl'' \leq sl_{2}$ and that the amount of super slots in the slot range from $sl'$ to $sl''$ is less than two times the amount of adversarial slots (indexed by the lottery $L_{1}$) in the same period of time.
\end{coqthm}
\begin{proof}
  Follows directly from \cref{thm:tcp} which can be instantiated with any lottery. 
\end{proof}
The way to interpret this theorem is that if a sufficiently long prefix of the lottery is drawn at random then we can get the guarantees that we need. Let us now turn our attention to what the concrete probability for a randomly drawn lottery is which is the theme for the next section. 
}%

\end{document}